\documentclass[a4paper,twocolumn,journal]{IEEEtran}
\usepackage[T1]{fontenc}
\usepackage{color}
\usepackage{array}
\usepackage{verbatim}
\usepackage{float}
\usepackage{textcomp}
\usepackage{amsmath}
\usepackage{amsthm}
\usepackage{amssymb}
\usepackage{graphicx}
\usepackage[unicode=true,
  bookmarks=true,bookmarksnumbered=true,bookmarksopen=true,bookmarksopenlevel=1,
  breaklinks=false,pdfborder={0 0 0},pdfborderstyle={},backref=false,colorlinks=false]
  {hyperref}
 \hypersetup{pdftitle={Your Title},
  pdfauthor={Your Name},
  pdfpagelayout=OneColumn, pdfnewwindow=true, pdfstartview=XYZ, plainpages=false}
  \usepackage{breakurl}

\makeatletter

\floatstyle{ruled}
\newfloat{algorithm}{tbp}{loa}
\providecommand{\algorithmname}{Algorithm}
\floatname{algorithm}{\protect\algorithmname}

\theoremstyle{plain}
\newtheorem{thm}{\protect\theoremname}
\theoremstyle{plain}
\newtheorem{prop}[thm]{\protect\propositionname}
\theoremstyle{definition}
\newtheorem{defn}[thm]{\protect\definitionname}
\theoremstyle{remark}
\newtheorem{rem}[thm]{\protect\remarkname}
\theoremstyle{plain}
\newtheorem{lem}[thm]{\protect\lemmaname}

\usepackage[caption=false,font=footnotesize]{subfig}
\usepackage{graphicx}
\usepackage[english]{babel}
\usepackage{times}
\usepackage{amsmath}
\usepackage{amstext}
\usepackage{amsfonts}
\usepackage{amssymb}
\usepackage{psfrag}
\usepackage{algorithmic}
\allowdisplaybreaks
\newcommand{\x}{l}

\newlength\myindent
    \usepackage{dsfont}
    
    \renewcommand{\mathbb}{\mathds}

\@ifundefined{showcaptionsetup}{}{%
 \PassOptionsToPackage{caption=false}{subfig}}
\usepackage{subfig}
\makeatother

\providecommand{\definitionname}{Definition}
\providecommand{\lemmaname}{Lemma}
\providecommand{\propositionname}{Proposition}
\providecommand{\remarkname}{Remark}
\providecommand{\theoremname}{Theorem}

\begin{document}

\title{Resource Optimization and Power Allocation in In-band Full~Duplex
(IBFD)-Enabled Non-Orthogonal Multiple Access Networks}

\author{Mohammed~S.~Elbamby,~\emph{Student~Member,~IEEE,} Mehdi~Bennis,~\emph{Senior~Member,~IEEE,}
Walid~Saad,~\emph{Senior~Member,~IEEE,} Mérouane~Debbah,~\emph{Fellow,~IEEE,}
and~Matti~Latva-aho,~\emph{Senior~Member,~IEEE}  \thanks{ Manuscript received January 27, 2017; revised April 21, 2017; Accepted May 22, 2017. This research has been supported by TEKES grant 2364/31/2014, the Academy of Finland  CARMA  project,   the U.S. National Science Foundation under Grant CNS-1460316, and the ERC Starting Grant 305123 MORE (Advanced Mathematical Tools for Complex Network Engineering).  The authors would like to thank Kien-Giang Nguyen and Trung Kien Vu for their constructive comments}\thanks{Mohammed~S.~Elbamby, Mehdi~Bennis, and Matti~Latva-aho are with
the{\small{} Centre for Wireless Communications, University of Oulu,
Finland, email: \{mohammed.elbamby,mehdi.bennis,matti.latva-aho\}@oulu.fi.
}Mehdi~Bennis is also with the Department of Computer Engineering,
Kyung Hee University, South Korea.}\thanks{Walid~Saad is with the{\small{} Wireless@VT, Bradley Department of
Electrical and Computer Engineering, Virginia Tech, Blacksburg, VA,
USA, email: walids@vt.edu}.}\thanks{Mérouane Debbah is with the{\small{} Large Networks and System Group
(LANEAS), CentraleSupélec, Université Paris-Saclay, Gif-sur-Yvette,
France}, and also with {\small{}the Mathematical and Algorithmic Sciences
Lab, Huawei France, Paris, France, email: merouane.debbah@huawei.com.}}}
\maketitle
\begin{abstract}
In this paper, the problem of uplink (UL) and downlink (DL) resource
optimization, mode selection and power allocation is studied for wireless
cellular networks under the assumption of in-band full duplex (IBFD)
base stations, non-orthogonal multiple access (NOMA) operation, and
queue stability constraints. The problem is formulated as a network
utility maximization problem for which a Lyapunov framework is used
to decompose it into two disjoint subproblems of auxiliary variable
selection and rate maximization. The latter is further decoupled into
a user association and mode selection (UAMS) problem and a UL/DL power
optimization (UDPO) problem that are solved concurrently. The UAMS
problem is modeled as a many-to-one matching problem to associate
users to small cell base stations (SBSs) and select transmission mode
(half/full-duplex and orthogonal/non-orthogonal multiple access),
and an algorithm is proposed to solve the problem converging to a
pairwise stable matching. Subsequently, the UDPO problem is formulated
as a sequence of convex problems and is solved using the concave-convex
procedure. Simulation results demonstrate the effectiveness of the
proposed scheme to allocate UL and DL power levels after dynamically
selecting the operating mode and the served users, under different
traffic intensity conditions, network density, and self-interference
cancellation capability. The proposed scheme is shown to achieve up
to 63\% and 73\% of gains in UL and DL packet throughput, and 21\%
and 17\% in UL and DL cell edge throughput, respectively, compared
to existing baseline schemes.
\end{abstract}


\section{Introduction and related work}

Due to the rapid increase in the demand for wireless data traffic,
there has been substantial recent efforts to develop new solutions
for improving the capacity of wireless cellular networks, by enhancing
the access to the scarce radio spectrum resources \cite{5g_cap_mag}.
In view of the scarcity of the frequency resources, smart duplexing,
resource allocation, and interference management schemes are crucial
to utilize the available resources in modern cellular networks and
satisfy the needs of their users.

\begin{figure}[htbp]
\includegraphics[width=1\columnwidth]{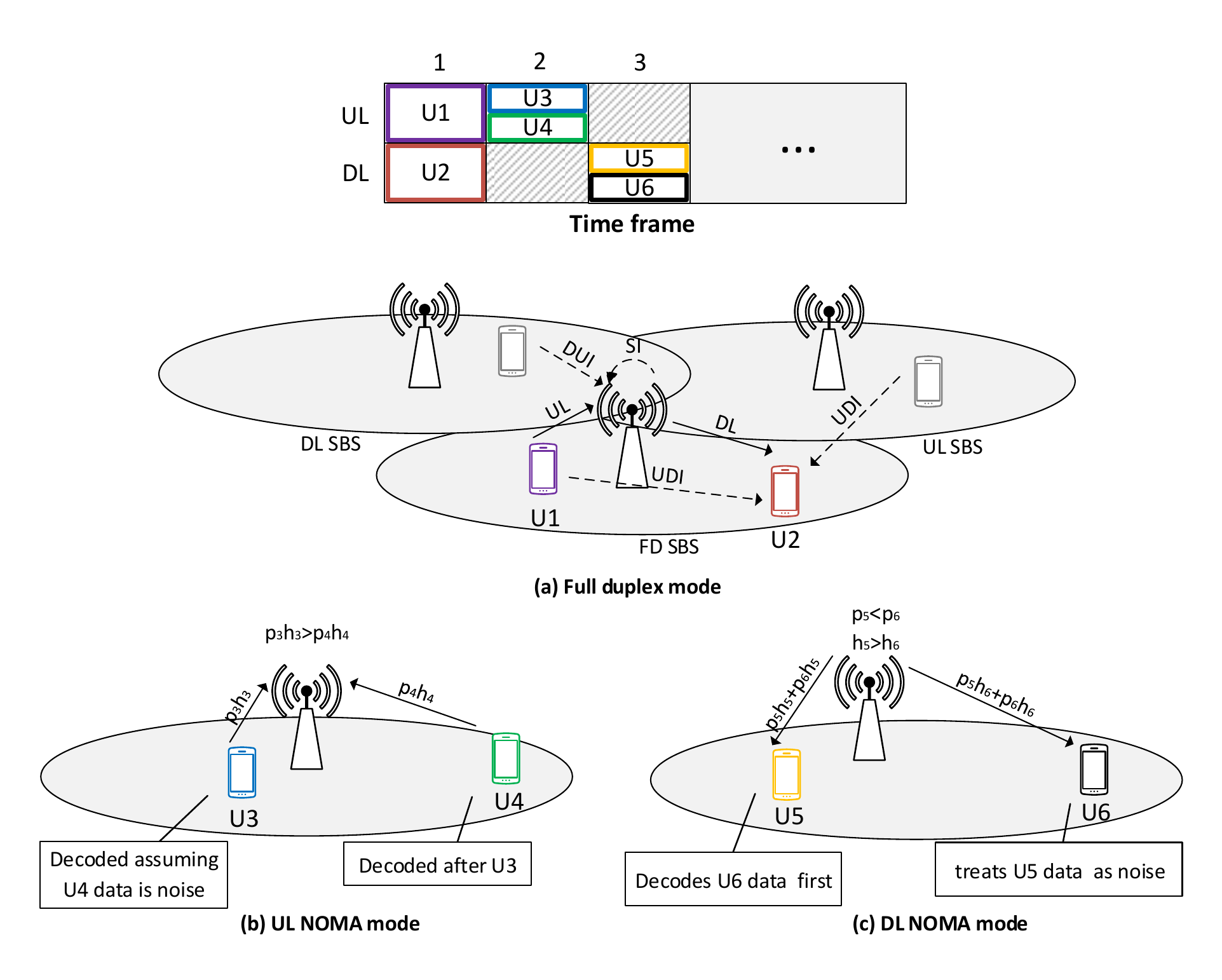}
\centering{}\caption{\label{fig:FD_NOMA}An illustration of (a) FD (b) UL NOMA and (c)
DL NOMA operations, UDI: UL-to-DL interference, DUI: DL-to-UL interference
and SI: self-interference.}
\end{figure}

In-band full duplex (IBFD) communication has the potential of doubling
the resource utilization of cellular networks through transmitting
in the uplink (UL) and downlink (DL) directions using the same time
and frequency resources. However, this comes at the cost of experiencing
higher levels of interference, not the least of which is the self-interference
(SI) leaked from the transmitter to the receiver of a IBFD radio.
Moreover, IBFD cellular networks suffer from additional interference
generated from having base stations and users transmitting on the
same channel. These interference types, namely, DL-to-UL interference
and UL-to-DL interference significantly degrade the performance of
IBFD networks. Figure \ref{fig:FD_NOMA} (a) illustrates desired and
interference signals in a FD-enabled base station.

Recent results have shown significant advances in the SI cancellation
of wireless IBFD radios, putting full duplex (FD) operation in wireless
cellular networks within reach. By employing a combination of analog
and digital cancellation techniques \cite{rice_SI,stanford_SI}, it
was shown that as much as $110$ dB of SI cancellation is achieved.
Another approach to avoid SI is to emulate a FD base station using
two spatially separated and coordinated half duplex (HD) base stations
\cite{petar_letter}. The performance of FD is investigated in \cite{Hirley_SI,hybrid_quek}
using tools from stochastic geometry. In \cite{Hirley_SI}, the combined
effect of SI and FD interference is analyzed. Throughput analysis
are conducted in \cite{hybrid_quek} with respect to network density,
interference and SI levels. A system-level simulation study of the
performance of IBFD ultra-dense small cell networks is conducted in
\cite{aalborg_simulation_FD}. The study concludes that FD may be
useful for asymmetric traffic conditions. However, none of these studies
take into account user selection and power allocation, which are key
factors to mitigate interference in a IBFD environment.

To further increase the utilization of frequency resources, the possibility
of multiple users sharing the same channel by exploiting differences
in the power domain has recently been proposed in the context of non-orthogonal
multiple access (NOMA) \cite{Noma_docomo}. Different from orthogonal
multiple access (OMA) schemes, NOMA allows multiple users to occupy
the same channel, whereas multi-user detection methods, such as successive
interference cancellation (SIC) are used to decode users' messages
\cite{NOMA_survey}. In DL NOMA, users with higher channel gains can
decode and cancel messages of users with lower channel gains before
decoding their own messages, whereas users with lowest channel gains
decode their messages first, treating other users' messages as noise.
In UL NOMA, base station performs successive decoding and cancellation
of different users data, ranked by their channel strength. The UL
and DL NOMA operation is illustrated for a two-user case in Figure
\ref{fig:FD_NOMA} (b) and (c), respectively. The decoding order as
well as the power allocation are key factors that affect the performance
of NOMA. Since NOMA comes with an additional receiver complexity for
the decoding phase, practical limitations on the maximum number of
users operating in NOMA must be considered. The combination of UL
NOMA and SIC has been shown to also enhance the cell-edge user throughput
in \cite{UL_NOMA_edge}.

User scheduling and power optimization are crucial factors in achieving
the promised gains of both FD and NOMA schemes. Intra-cell and inter-cell
cross-link interference occur due to IBFD operation. Therefore, smart
resource and interference management schemes are needed to reap the
gains of FD and NOMA operation.  In \cite{goyal_single_FD}, a hybrid
HD/FD scheduler is proposed that can assign FD resources only when
it is advantageous over HD resource assignment. An auction-based algorithm
to pair UL and DL users in a single-cell IBFD network is proposed
in \cite{Ericsson_FD_single}. User scheduling and power allocation
algorithms are presented for IBFD networks in \cite{zorzi_imperfect}
and for DL NOMA networks in \cite{NOMA_matching_conf,parida_NOMA,fang_NOMA}.
In the proposed approach of \cite{NOMA_matching_conf}, a many-to-many
matching algorithm is used to assign channel to users in a single
cell scenario. The performance of NOMA in the UL is investigated in
\cite{Rahim_NOMA} using an iterative channel allocation algorithm.
All of the above works focus on single-cell optimization and hence,
avoid handling the inter-cell interference. A distributed sub-optimal
algorithm for the problem of user selection and power allocation in
a multi-cell full duplex network discussed in \cite{goyal_multi_FD}
showing that IBFD can achieve up to double the throughput of HD in
indoor scenarios and $65\%$ in outdoor scenarios. However, the proposed
algorithm relies on information exchange between neighboring base
stations. It also does not take into account queue state information
(QSI), which has a significant impact on the user packet throughput.
Several algorithms are proposed in \cite{zhu_radwa_mode,boya_matching_poster,marios_FD}
to optimize user pairing/scheduling, and power allocation. These algorithms
are either centralized or inter-cell interference is overlooked. In
\cite{Ekram_NOMA_clustering}, the authors consider a multi-cell UL
and DL NOMA system where they propose a user grouping and power optimization
framework. The optimal power allocation is derived for a single macro
cell and limited number of users. The problem of UL/DL decoupled user
association is studied in \cite{Ekram_FD_dude_J} for multi-tier IBFD
networks where many-to-one matching algorithms are proposed to solve
the problem. However, dynamic user scheduling and power allocation
are not considered in the optimization problem. Studying the operation
of both FD and NOMA was first considered in \cite{Sun_FD_NOMA}, where
the problem of subcarrier and power allocation in a multicarrier and
single-cell FD-NOMA system is optimized. Taking into account the user
traffic dynamic in the optimization problem is also missing in the
literature. A summary of the state of the art contributions to FD
and NOMA resource and power optimization problems is provided in Table~\ref{tab:Summary-of-contributions}.

\subsection{Paper Contribution}

The main goal of this paper is to study the problem of user association
and power allocation in a multi-cell IBFD-enabled network operating
in NOMA. Our objective is to investigate the benefits of operating
in HD or FD, as well as in OMA or NOMA modes, depending on traffic
conditions, network density, and self-interference cancellation capabilities.
We propose a Lyapunov based framework to jointly optimize the duplex
mode selection, user association and power allocation. The problem
is formulated as a network utility maximization where the main objective
is to minimize a Lyapunov drift-plus-penalty function that strikes
a balance between maximizing a system utility that is a function of
the data rate, and stabilize the system traffic queues. The optimization
problem is decomposed into two subproblems that are solved independently
per SBS. User association and mode selection (UAMS) is formulated
as a many-to-one matching problem. A distributed matching algorithm
aided by an inter-cell interference learning mechanism is proposed
which is shown to converge to a pairwise stable matching. The matching
algorithm allows SBSs to switch their operation between HD and FD
and select either OMA or NOMA schemes to serve its users. Following
that, UL/DL power optimization (UDPO) is formulated as a sequence
of convex problems and an iterative algorithm to allocate the optimal
power levels for the matched users and their SBSs is proposed. %

{} Simulation results show that using the proposed matching algorithm,
a network can dynamically select when to operate in HD or FD and when
to use OMA or NOMA to serve different users. It is also shown that
significant gains in UL and DL user throughput and packet throughput
can be achieved using the proposed scheme, as compared to different
baseline schemes. %

The rest of this paper is organized as follows: Section~\ref{sec:sys_mod}
describes the system model. The optimization problem is formulated
and decomposed into disjoint subproblems using the Lyapunov framework
in Section~\ref{sec:ProblemForm}. Section~\ref{sec:rate_power_prob}
discusses the proposed matching algorithm and power allocation scheme.
The performance of the proposed framework is analyzed in Section~\ref{sec:sim_res}.
Finally, Section~\ref{sec:conc} concludes the paper and outlines
future open research lines.

\emph{Notations:} Lower case symbols represent scalars, boldface symbols
represent vectors, whereas upper case calligraphic symbols represent
sets. $\mathbb{1}_{\{condition\}}$ is the indicator function, which
equals to $1$ whenever $condition$ is true and $0$ otherwise, $|\mathcal{X}|$
denotes the cardinality of a set $\mathcal{X}$, ${[x]}^{+}=\max\{x,0\}$, and $\nabla f(\boldsymbol{x})$ is the gradient of the function
$f(\boldsymbol{x})$.

\begin{table*}[tbh]
\begin{centering}
\caption{Summary of existing literature in FD and NOMA-based resource allocation
problems.\label{tab:Summary-of-contributions}}
\par\end{centering}
\begin{centering}
\subfloat[FD]{
\centering{}%
\begin{tabular}{|c|>{\centering}p{2.7cm}|>{\centering}p{2cm}|>{\centering}p{4.6cm}|>{\centering}p{1.5cm}|>{\centering}p{1.5cm}|}
\hline
\textbf{Reference} & \textbf{Network scenario} & \textbf{Implementation} & \textbf{FD scheduling} & \textbf{Power allocation} & \textbf{Queue dynamics}\tabularnewline
\hline
\hline
\cite{goyal_single_FD} & single-cell & local & HD/FD mode selection & $\times$ & $\times$\tabularnewline
\hline
\cite{Ericsson_FD_single} & single-cell & local & FD user pairing and channel allocation & $\surd$ & $\times$\tabularnewline
\hline
\cite{zorzi_imperfect} & single-cell & local & OFDMA channel allocation & $\surd$ & $\times$\tabularnewline
\hline
\cite{goyal_multi_FD} & multi-cell & local  & suboptimal HD/FD user selection & $\surd$ & $\times$\tabularnewline
\hline
\cite{zhu_radwa_mode} & multi-cell (subcarrier is reused once) & central & mode selection and subcarrier allocation & $\surd$ & $\times$\tabularnewline
\hline
\cite{boya_matching_poster} & multi-cell & central & matching subcarriers to user pairs & $\surd$ & $\times$\tabularnewline
\hline
\cite{marios_FD} & multi-cell & local & local scheduling, ignores inter-cell interference & $\surd$ & $\times$\tabularnewline
\hline
\cite{Ekram_FD_dude_J} & multi-cell & central/local & UL/DL decoupled user association & $\times$ & $\times$\tabularnewline
\hline
\end{tabular}}
\par\end{centering}
\centering{}\subfloat[NOMA]{
\centering{}%
\begin{tabular}{|c|c|c|>{\centering}p{6cm}|>{\centering}p{1.5cm}|}
\hline
\textbf{Reference} & \textbf{Link scenario} & \textbf{Network scenario} & \textbf{Scheduling and power allocation scheme} & \textbf{Queue dynamics}\tabularnewline
\hline
\hline
 \cite{NOMA_matching_conf} & DL & single-cell  & matching algorithm & $\times$\tabularnewline
\hline
 \cite{Rahim_NOMA}  & UL & single-cell  & iterative subcarrier and power allocation & $\times$\tabularnewline
\hline
\cite{fang_NOMA} & DL & single-cell  & subchannel assignment and power allocation & $\times$\tabularnewline
\hline
\cite{parida_NOMA} & DL & single-cell & user selection and power optimization & $\times$\tabularnewline
\hline
\cite{Ekram_NOMA_clustering} & UL + DL & multi-cell  & optimal power allocation \\
for limited number of users & $\times$\tabularnewline
\hline
\cite{Boya_NOMA_jou} & DL & multi-cell & matching algorithm and power allocation & $\times$\tabularnewline
\hline
\cite{NOMA_UL_GC} & UL & heterogeneous  & user clustering and power allocation  & $\times$\tabularnewline
\hline
\cite{Sun_FD_NOMA} & UL+DL+FD & single-cell  & joint subchannel and power allocation & $\times$\tabularnewline
\hline
\end{tabular}}
\end{table*}

\section{System Model\label{sec:sys_mod}}

We consider a time division duplexing (TDD) system in a network of
small cells, where a set $\mathcal{B}$ of $B$ SBSs operate in either
FD or HD mode, with SI cancellation capability of $\zeta$, whereas
a set \textbf{$\mathcal{U}$} of $U$ users are available in the network
and are restricted to HD operation. An open-access policy is considered
where users associate to any SBS in the UL or DL. Furthermore, we
assume that base stations and users can operate in NOMA scheme, where
multiple users can share the same time-frequency resource in either
UL or DL. To cancel the resulting multi-user interference, SBSs or
users operating in NOMA can perform SIC at the receiver side. In this
regard, we assume that the decoding ordering is done in an descending
order of channel strength in DL NOMA. Moreover, we adopt the ascending
order of channel strength in UL NOMA, which was analytically shown
in \cite{NOMA_UL_GC} to outperform OMA.

UL and DL service rates of user $u$ are defined as:
\begin{eqnarray}
r_{u}^{\text{UL}}(t) & = & \sum_{b\in\mathcal{B}}x_{bu}^{\text{UL}}(t)R_{bu}^{\text{UL}}(t),\nonumber \\
 & = & \sum_{b\in\mathcal{B}}x_{bu}^{\text{UL}}(t)f_{b}\log_{2}(1+\textrm{\ensuremath{\Gamma}}_{bu}^{\textrm{UL}}(t))\label{eq:UL_rate}
\end{eqnarray}
\begin{eqnarray}
r_{u}^{\text{DL}}(t) & = & \sum_{b\in\mathcal{B}}x_{bu}^{\text{DL}}(t)R_{u}^{\text{DL}}(t),\nonumber \\
 & = & \sum_{b\in\mathcal{B}}x_{bu}^{\text{DL}}(t)f_{b}\log_{2}(1+\textrm{\ensuremath{\Gamma}}_{bu}^{\textrm{DL}}(t))\label{eq:DL_rate}
\end{eqnarray}
where $x_{bu}^{\text{UL}}(t)\in X^{\text{UL}}$, $x_{bu}^{\text{DL}}(t)\in X^{\text{DL}}$
are indicator variables that user $u$ is associated to SBS $b$ at
time instant $t$ in UL or DL, respectively, $f_{b}$ is the allocated
bandwidth, and $R_{bu}^{\text{UL}}$ and $R_{bu}^{\text{DL}}$ are
the UL and DL data rates between SBS $b$ and user $r$. The UL and
DL signal to interference plus noise ratios between SBS $b$ and user
$u$ at time instant $t$, are given by:

\begin{align}
\textrm{\ensuremath{\Gamma}}_{bu}^{\textrm{UL}}(t)\negthinspace & =\negthinspace\frac{p_{u}^{\text{UL}}(t)h_{bu}(t)}{N_{0}\negthinspace+\negthinspace I_{b}^{\textrm{UL-UL}}(t)\negthinspace+\negthinspace I_{b}^{\textrm{DL-UL}}(t)\negthinspace+\negthinspace I_{bu}^{\textrm{NOMA-UL}}(t)\negthinspace+\negthinspace p_{b}^{\text{DL}}(t)\negthinspace/\negthinspace\zeta},\label{eq:SINR_UL}\\
\textrm{\ensuremath{\Gamma}}_{bu}^{\textrm{DL}}(t) & \negthinspace=\negthinspace\frac{p_{bu}^{\text{DL}}(t)h_{bu}(t)}{N_{0}+I_{u}^{\textrm{DL-DL}}(t)+I_{u}^{\textrm{UL-DL}}(t)+I_{bu}^{\textrm{NOMA-DL}}(t)},\label{eq:SINR_DL}
\end{align}
where $p_{u}^{\text{UL}}$ is the UL transmit power of user $u$,
and $p_{b}^{\text{DL}}=\sum_{u\in\mathcal{U}}p_{bu}^{\text{DL}}$
is the total DL transmit power of SBS $b$, $h_{x,y}(t)=|g_{x,y}(t)|^{2}$
is the channel gain between the two nodes $x$ and $y$, $g_{x,y}(t)$
is the propagation channel between the two nodes $x$ and $y$, and
the interference terms in (\ref{eq:SINR_UL}) and (\ref{eq:SINR_DL})
are expressed as follows\footnote{Note that the term $I_{\textrm{UL-DL}}$ includes the intra-cell interference,
to account for the interference due to FD operation. }:
\begin{alignat*}{1}
I_{b}^{\textrm{UL-UL}}(t) & =\sum_{u'\in\mathcal{U}\backslash\{u\}}p_{u'}^{\text{UL}}(t)h_{bu'}(t),\\
I_{b}^{\textrm{DL-UL}}(t) & =\sum_{b'\in\mathcal{B}\backslash\{b\}}p_{b'}^{\text{DL}}(t)h_{b'b}(t),\\
I_{u}^{\textrm{DL-DL}}(t) & =\sum_{b'\in\mathcal{B}\backslash\{b\}}p_{b'}^{\text{DL}}(t)h_{b'u}(t),\\
I_{u}^{\textrm{UL-DL}}(t) & =\sum_{u'\in\mathcal{U}}p_{u'}^{\text{UL}}(t)h_{u',u}(t),\\
I_{bu}^{\textrm{NOMA-UL}}(t) & =\sum_{\substack{u':x_{bu'}^{\text{UL}}=1,\\
h_{bu'}<h_{bu}
}
}p_{u'}^{\text{UL}}(t)h_{bu'}(t),\\
I_{bu}^{\textrm{NOMA-DL}}(t) & =\sum_{\substack{u':x_{bu'}^{\text{DL}}=1,\\
h_{bu'}>h_{bu}
}
}p_{bu'}^{\text{DL}}(t)h_{bu}(t),
\end{alignat*}
 $p_{b}^{\text{DL}}(t)/\zeta$ is the leaked SI and $N_{0}$ is the
noise variance. To guarantee a successful NOMA operation in the DL,
a user $u'$ should decode the data of user $u$ with an SINR level
$\textrm{\ensuremath{\Gamma}}_{bu}^{u'\textrm{DL}}(t)$ that is at
least equal to the user $u$ received SINR $\textrm{\ensuremath{\Gamma}}_{bu}^{\textrm{DL}}(t)$.
Otherwise, the data rate of user $u$ is higher than what user $u'$
can decode. Accordingly, the inequality $\textrm{\ensuremath{\Gamma}}_{bu}^{u'\textrm{DL}}(t)\geq\textrm{\ensuremath{\Gamma}}_{bu}^{\textrm{DL}}(t)$
must hold, where:
\begin{multline*}
\textrm{\ensuremath{\Gamma}}_{bu}^{u'\textrm{DL}}(t)=\\
\qquad\frac{p_{bu}^{\text{DL}}(t)h_{bu'}(t)}{p_{bu'}^{\text{DL}}(t)h_{bu'}(t)\negthinspace+\negthinspace N_{0}\negthinspace+\negthinspace I_{u'}^{\textrm{DL-DL}}(t)\negthinspace+\negthinspace I_{u'}^{\textrm{UL-DL}}(t)\negthinspace+\negthinspace I_{bu'}^{\textrm{NOMA-DL}}(t)}.
\end{multline*}

Let $\boldsymbol{Q}^{\text{UL}}(t)=[Q_{1}^{\text{UL}}(t),...,Q_{U}^{\text{UL}}(t)]$,
and $\boldsymbol{Q}^{\text{DL}}(t)=[Q_{1}^{\text{DL}}(t),...,Q_{U}^{\text{DL}}(t)]$
denote the UL and DL traffic queues at time instant $t$, with $Q_{u}^{\x}(t),\x\in\{\text{UL},\text{DL}\}$
representing the UL/DL queues of a given user $u$. Then, the queue
dynamics for user $u$ are given by:

\begin{equation}
Q_{u}^{\x}(t+1)={[Q_{u}^{\x}(t)-r_{u}^{\x}(t)]}^{+}+A_{u}^{\x}(t),\label{eq:traffic_Q_UL_DL}
\end{equation}
where $A_{u}^{\x}$ is the data arrival for user $u$ in the link
direction $\x$ at time instant $t$, which is assumed to be independent
and identically distributed with mean $\lambda_{u}^{\x}/\mu_{u}^{\x}>0$,
and bounded above by the finite value $A_{\max}^{\x}$. $\lambda_{u}^{\x}$
and $1/\mu_{u}^{\x}$ are the mean packet arrival rate and mean packet
size, which follow Poisson and exponential distributions, respectively.
Here ${[.]}^{+}$ indicates that the actual served rate cannot exceed
the amount of traffic in a queue. To satisfy queue stability requirements,
SBSs need to ensure that traffic queues are mean rate stable. This
is equivalent to ensuring that the average service rate is higher
or equal to the average data arrival, i.e., $\bar{A}_{u}^{\x}\leq\bar{r}_{u}^{\x}$.

\section{Problem Formulation\label{sec:ProblemForm}}

In small cell networks, the network utility is affected by various
factors such as user association, UL/DL mode selection, OMA/NOMA mode
selection, and UL and DL power levels. Let $X^{\text{UL}}=[x_{bu}^{\text{UL}}]$
and $X^{\text{DL}}=[x_{bu}^{\text{DL}}]$ be the UL and DL user association
matrices, $\boldsymbol{p}^{\text{UL}}=[p_{1}^{\text{UL}},...,p_{U}^{\text{UL}}]$
be the UL power vector, and $\boldsymbol{p}_{b}^{\text{DL}}=[p_{b1}^{\text{DL}},...,p_{bU}^{\text{DL}}]$
be the DL power vector of SBS $b$. Therefore, a joint optimization
problem of user association, mode selection and power allocation to
maximize a continuous utility function of time-averaged UL and DL
service rates, is cast as follows:

\begin{subequations}\label{eq:objective}
\begin{align}
\max_{\substack{X^{\text{UL}},X^{\text{DL}},\\
\boldsymbol{p}^{\text{UL}},\{\boldsymbol{p}_{b}^{\text{DL}}\}
}
}\quad & U\bigl(\{\bar{r}_{u}^{\text{UL}}\},\{\bar{r}_{u}^{\text{DL}}\}\bigr)\\
 & =\sum_{u\in\mathcal{U}}\Bigl(f_{^{\text{UL}}}(\bar{r}_{u}^{\text{UL}})+f_{^{\text{DL}}}(\bar{r}_{u}^{\text{DL}})\Bigr)\nonumber \\
\text{subject to}\quad & \bar{A}_{u}^{\text{UL}}\leq\bar{r}_{u}^{\text{UL}},\;\bar{A}_{u}^{\text{DL}}\leq\bar{r}_{u}^{\text{DL}},\;\forall u\in\mathcal{U},\label{eq:const1}\\
 & \bar{p}_{u}^{\text{UL}}\leq\bar{\delta}_{u}^{\text{UL}},\;\bar{p}_{b}^{\text{DL}}\leq\bar{\delta}_{b}^{\text{DL}},\;\forall u\in\mathcal{U},\label{eq:const3}\\
 & \delta_{u}^{\text{UL}}\leq P_{\max}^{\text{UL}},\;\forall u\in\mathcal{U},\label{eq:const4}\\
 & \delta_{b}^{\text{DL}}\leq P_{\max}^{\text{DL}},\;\forall b\in\mathcal{B},\label{eq:const5}\\
 & \sum_{b\in\mathcal{B}}x_{bu}^{\text{UL}}+x_{bu}^{\text{DL}}\leq1,\;\forall u\in\mathcal{U},\label{eq:const6}\\
 & \Bigl(\sum_{u\in\mathcal{U}}x_{bu}^{\text{UL}}\Bigr).\Bigl(\sum_{u\in\mathcal{U}}x_{bu}^{\text{DL}}\Bigr)\leq1,\;\forall b\in\mathcal{B}.\label{eq:const7}\\
 & \sum_{u\in\mathcal{U}}x_{bu}^{\x}\leq q,\;\forall b\in\mathcal{B},\;\forall\x\in\{\text{UL},\text{DL}\}.\label{eq:const8}\\
 &   Y_{uu'}\geq0\;\;\forall u,u'\in\mathcal{U}, \label{eq:const_feas}
\end{align}
\end{subequations} Constraint (\ref{eq:const1}) ensures the stability
of UL and DL traffic queues. Constraints (\ref{eq:const3}) limit
the effect of aggregate interference by maintaining the average transmit
powers of SBSs and users constrained by threshold values $\delta_{b}^{\text{UL}}$
and $\delta_{u}^{\text{DL}}$. Here, $P_{\max}^{\text{UL}}$ and $P_{\max}^{\text{DL}}$
denote the maximum UL and DL powers. Note that maintaining the average
transmit power below a threshold value allows SBSs to limit their
leaked interference within pre-defined limits without the need to
exchange information with others, which, in turn, allows for distributed
implementation. Constraint (\ref{eq:const_feas}), where $Y_{uu'}=\mathbb{1}_{h_{bu'}>h_{bu}}x_{bu}^{\text{DL}}x_{bu'}^{\text{DL}}(\textrm{\ensuremath{\Gamma}}_{bu}^{u'\textrm{DL}}-\textrm{\ensuremath{\Gamma}}_{bu}^{\textrm{DL}})$
is to guarantee a successful SIC in the DL. Constraint (\ref{eq:const6})
states that a user is associated to a single SBS at a time in either
UL or DL. (\ref{eq:const7}) ensures that an SBS cannot operate in
both FD and NOMA at the same time instant, i.e., if it serves one
user in one direction it can serve at most one user in the other direction.
Finally, the number of users an SBS can simultaneously serve in UL
or DL NOMA is limited to a quota of $q$ users by constraint (\ref{eq:const8})\footnote{Note that, in theory, the number of users that can be served simulateaneously
using NOMA is unrestricted. However, we impose a quota to avoid high
complexity SIC in the receiver side if a high number of users are
scheduled. }. We note that the restriction in (\ref{eq:const7}) is imposed to
avoid  high intra-cell interference resulting from operating in both
FD and NOMA, in which case assuming that users can perform SIC of
multiple users transmitting in both directions might be impractical
as the number of users scheduled under NOMA increases.

\sloppy We define sets of UL and DL auxiliary variables for each
user, denoted as $\{\gamma_{u}^{\text{UL}}\},\{\gamma_{u}^{\text{DL}}\},\forall u\in U_{b}$
such that the problem in (\ref{eq:objective}) is transformed from
a utility function of time-averaged rates into an equivalent optimization
problem with a time-averaged utility function of instantaneous rates.
Subsequently, the problem can be solved in an online manner at each
time instant without prior knowledge of the statistical information
of the network variables. In this regard, we define the equivalent
optimization problem as follows:

\begin{subequations}\label{eq:transformed_objective}
\begin{align}
\max_{\substack{X_{b}^{\text{UL}},X_{b}^{\text{DL}},\boldsymbol{p}^{\text{UL}},\{\boldsymbol{p}_{b}^{\text{DL}}\}\\
\{\gamma_{u}^{\text{UL}}\},\{\gamma_{u}^{\text{DL}}\}
}
} & \overline{U(\{\gamma_{u}^{\text{UL}}\},\{\gamma_{u}^{\text{DL}}\})}\quad\\
 & =\overline{\sum_{u\in\mathcal{U}}\Bigl(f_{^{\text{UL}}}(\gamma_{u}^{\text{UL}})+f_{^{\text{DL}}}(\gamma_{u}^{\text{DL}})\Bigr)}\nonumber \\
\text{subject to}\quad & \bar{\gamma}_{u}^{\text{UL}}\leq\bar{r}_{u}^{\text{UL}},\;\bar{\gamma}_{u}^{\text{DL}}\leq\bar{r}_{u}^{\text{DL}},\;\forall u\in\mathcal{U},\label{eq:aux_const}\\
 & \gamma_{u}^{\text{UL}}\leq r_{\max}^{\text{UL}},\;\gamma_{u}^{\text{DL}}\leq r_{\max}^{\text{DL}},\;\forall u\in\mathcal{U},\\
 & (\ref{eq:const1})-(\ref{eq:const_feas}).
\end{align}
\end{subequations}

The transformed problem in (\ref{eq:transformed_objective}) is equivalent
to the original problem in (\ref{eq:objective}). To show this \cite{neely_lyap_2},
suppose that $U_{1}^{*}$ and $U_{2}^{*}$ are the optimal utility
values of problems (\ref{eq:objective}) and (\ref{eq:transformed_objective}),
and $X_{b}^{*\text{UL}},X_{b}^{*\text{DL}},\boldsymbol{p}^{*\text{UL}},\{\boldsymbol{p}_{b}^{*\text{DL}}\}$
are the optimal stationary and randomized configurations that solve
(\ref{eq:objective}) and correspond  to users' optimal expected rates
$\{\bar{r}_{u}^{*\text{UL}}\},\{\bar{r}_{u}^{*\text{DL}}\}$. Then,
$U_{1}^{*}=U\bigl(\{\bar{r}_{u}^{*\text{UL}}\},\{\bar{r}_{u}^{*\text{DL}}\}\bigr)$.

At all time instants, the values $X_{b}^{*\text{UL}},X_{b}^{*\text{DL}},\boldsymbol{p}^{*\text{UL}},\{\boldsymbol{p}_{b}^{*\text{DL}}\}$
as well as the auxiliary variables $\{\gamma_{u}^{\text{UL}}=\bar{r}_{u}^{*\text{UL}}\},\{\gamma_{u}^{\text{DL}}=\bar{r}_{u}^{*\text{DL}}\}$
satisfy the constraints of problem (\ref{eq:transformed_objective})
and correspond to the utility value $\overline{U(\{\gamma_{u}^{\text{UL}}\},\{\gamma_{u}^{\text{DL}}\})}=U_{1}^{*}$.
As this value is not necessarily optimal in problem (\ref{eq:transformed_objective}),
then $U_{2}^{*}\geq U_{1}^{*}$.

Then, let $X_{b}^{\text{UL}},X_{b}^{\text{DL}},\boldsymbol{p}^{\text{UL}},\{\boldsymbol{p}_{b}^{\text{DL}}\}\{\gamma_{u}^{\text{UL}}\},\{\gamma_{u}^{\text{DL}}\}$
be the configurations that solve the transformed problem (\ref{eq:transformed_objective}).
These values satisfy the original problem (\ref{eq:objective}) and
correspond to a utility value that is not greater than $U_{1}^{*}.$
Therefore:
\begin{align*}
U_{1}^{*} & \geq U\bigl(\{\bar{r}_{u}^{\text{UL}}\},\{\bar{r}_{u}^{\text{DL}}\}\bigr)\\
 & \geq U\bigl(\{\bar{\gamma}_{u}^{\text{UL}}\},\{\bar{\gamma}_{u}^{\text{DL}}\}\bigr)\\
 & \geq\overline{U(\{\gamma_{u}^{\text{UL}}\},\{\gamma_{u}^{\text{DL}}\})}\\
 & =U_{2}^{*},
\end{align*}
where the second inequality is due to the constraint (\ref{eq:aux_const})
and the continuity of the utility function, the third inequality follows
from Jensen's inequality, and the last step is due to the configurations
being the optimal solution for (\ref{eq:transformed_objective}).
Therefore, one gets that $U_{2}^{*}=U_{1}^{*}$ and any solution to
(\ref{eq:transformed_objective}) is also a solution to (\ref{eq:objective}).

Next, we invoke the framework of Lyapunov optimization \cite{neely_lyapunov},
the time-average inequality constraints can be satisfied by converting
them into \emph{virtual} queues and maintaining their stability. Therefore,
we define the following virtual queues $\boldsymbol{H}^{\text{UL}}(t)=[H_{1}^{\text{UL}}(t),...,H_{U}^{\text{UL}}(t)]$,
$\boldsymbol{H}^{\text{DL}}(t)=[H_{1}^{\text{DL}}(t),...,H_{U}^{\text{DL}}(t)]$,
$\boldsymbol{Z}^{\text{UL}}(t)=[Z_{1}^{\text{UL}}(t),...,Z_{U}^{\text{UL}}(t)]$,
and $\boldsymbol{Z}^{\text{DL}}(t)=[Z_{1}^{\text{DL}}(t),...,Z_{b}^{\text{DL}}(t)]$
that correspond to the constraints over the auxiliary variables and
transmit power levels, respectively. Accordingly, the virtual queues
are updated as follows:
\begin{equation}
H_{u}^{\x}(t+1)={[H_{u}^{\x}(t)-r_{u}^{\x}(t)]}^{+}+\gamma_{u}^{\x}(t),\label{eq:aux_Q_UL_DL}
\end{equation}
\begin{equation}
Z_{u}^{\text{UL}}(t+1)={[Z_{u}^{\text{UL}}(t)-\delta_{u}^{\text{UL}}(t)]}^{+}+p_{u}^{\text{UL}}(t),\label{eq:power_Q_UL}
\end{equation}
\begin{equation}
Z_{b}^{\text{DL}}(t+1)={[Z_{b}^{\text{DL}}(t)-\delta_{b}^{\text{DL}}(t)]}^{+}+p_{b}^{\text{DL}}(t).\label{eq:power_Q_DL}
\end{equation}

Let a combination of the queues be $y(t)\triangleq[\boldsymbol{Q}^{\text{UL}}(t),\boldsymbol{Q}^{\text{DL}}(t),\boldsymbol{H}^{\text{UL}}(t),\boldsymbol{H}^{\text{DL}}(t),\boldsymbol{Z}^{\text{UL}}(t),\boldsymbol{Z}^{\text{DL}}(t)]$,
we define the Lyapunov function $L\bigl(y(t)\bigr)$ and the drift-plus-penalty
function $\Delta\bigl(y(t)\bigr)$ as:
\begin{multline}
\negthinspace L\bigl(y(t)\bigr)=\frac{1}{2}\Bigl\{\left\Vert \boldsymbol{Q}^{\text{UL}}(t)\right\Vert ^{2}+\left\Vert \boldsymbol{Q}^{\text{DL}}(t)\right\Vert ^{2}\\
\negthinspace+\negthinspace\left\Vert \boldsymbol{H}^{\text{UL}}(t)\right\Vert ^{2}\negthinspace+\negthinspace\left\Vert \boldsymbol{H}^{\text{DL}}(t)\right\Vert ^{2}\negthinspace+\negthinspace\left\Vert \boldsymbol{Z}^{\text{UL}}(t)\right\Vert ^{2}\negthinspace+\negthinspace\left\Vert \boldsymbol{Z}^{\text{DL}}(t)\right\Vert ^{2}\Bigr\},
\end{multline}
\begin{equation}
\Delta\bigl(y(t)\bigr)\negthinspace=\negthinspace\mathbb{E}\Bigl\{ L\bigl(y(t+1)\bigr)\negthinspace-\negthinspace L\bigl(y(t)\bigr)\negthinspace-\negthinspace vU(\{\gamma_{u}^{\text{UL}}\},\negthinspace\{\gamma_{u}^{\text{DL}}\})|Y(t)\Bigr\},\label{eq:drift_eq}
\end{equation}
where $v$ is a non-negative parameter that controls the tradeoff
between achieving the optimal utility and ensuring the stability of
the queues.
\begin{prop}
\label{prop:DPP}At time instant $t$, the Lyapunov drift-plus-penalty
satisfies the inequality in (\ref{eq:drift_plus_penalty_bound}) under
any control strategy and queue state, where $C=\frac{1}{2}\sum_{u\in\mathcal{U}}\Bigl[{A_{\max}^{\text{UL}}}^{2}+{A_{\max}^{\text{DL}}}^{2}+2{P_{\max}^{\text{UL}}}^{2}+3{r_{\max}^{\text{UL}}}^{2}+3{r_{\max}^{\text{DL}}}^{2}\Bigr]+\sum_{b\in\mathcal{B}}2{P_{\max}^{\text{DL}}}^{2}$
is a finite parameter.

\begin{figure*}[!t] 
\normalsize 

\begin{align}
\Delta\bigl(Y(t)\bigr) & \leq C\label{eq:drift_plus_penalty_bound}\\
 & +\stackrel{\textrm{fixed at time \ensuremath{t}}}{\overbrace{\mathbb{E}\{\sum_{u\in\mathcal{U}}\Bigl[(Q_{u}^{\text{UL}}(t))A_{u}^{\text{UL}}(t)+(Q_{u}^{\text{DL}}(t))A_{u}^{\text{DL}}(t)\Bigr]\}}}\nonumber \\
 & -\biggr[\stackrel{\textrm{impact of auxiliary variables and virtual queues }}{\overbrace{\mathbb{E}\{\sum_{u\in\mathcal{U}}\Bigl[v.f_{^{\text{UL}}}(\gamma_{u}^{\text{UL}}(t))-H_{u}^{\text{UL}}(t)\gamma_{u}^{\text{UL}}(t)+v.f_{^{\text{DL}}}(\gamma_{u}^{\text{DL}}(t))-H_{u}^{\text{DL}}(t)\gamma_{u}^{\text{DL}}(t)\Bigr]\}}\biggl]_{\#1}}\nonumber \\
 & -\biggr[\stackrel{\textrm{impact of scheduling and QSI }}{\overbrace{\mathbb{E}\{\sum_{u\in\mathcal{U}}\Bigl[(Q_{u}^{\text{UL}}(t)+H_{u}^{\text{UL}}(t))r_{u}^{\text{UL}}(t)+(Q_{u}^{\text{DL}}(t)+H_{u}^{\text{DL}}(t))r_{u}^{\text{DL}}(t)}}\nonumber \\
 & \qquad\qquad+\stackrel{\textrm{impact of power allocation and power queues}}{\overbrace{Z_{u}^{\text{UL}}(t)(\delta_{u}^{\text{UL}}(t)-p_{u}^{\text{UL}}(t))\Bigr]\}-\mathbb{E}\{\sum_{b\in\mathcal{B}}\Bigl[Z_{b}^{\text{DL}}(t)(\delta_{b}^{\text{DL}}(t)-p_{b}^{\text{DL}}(t))\Bigr]\}}\biggl]_{\#2}}\nonumber
\end{align}
\hrulefill 
\end{figure*}
\end{prop}
\begin{IEEEproof}
By squaring equations (\ref{eq:traffic_Q_UL_DL}), (\ref{eq:aux_Q_UL_DL}),
(\ref{eq:power_Q_UL}) and (\ref{eq:power_Q_DL}), and from the fact
that $({[a-b]}^{+}+c)^{2}\leq a^{2}+b^{2}+c^{2}-2a(b\text{\textminus}c),\forall a,b,c\geq0$,
we can show that: %

\[
{Q_{u}^{\x}}^{2}(t+1)-{Q_{u}^{\x}}^{2}(t)\negthinspace\leq\negthinspace{A_{u}^{\x}}^{2}(t)+{r_{u}^{\x}}^{2}(t)-2{Q_{u}^{\x}}(t)\negthinspace\Bigl(r_{u}^{\x}(t)-A_{u}^{\x}(t)\Bigr),
\]
\[
{H_{u}^{\x}}^{2}(t+1)-{H_{u}^{\x}}^{2}(t)\negthinspace\leq\negthinspace{\gamma_{u}^{\x}}^{2}(t)+{r_{u}^{\x}}^{2}(t)-2{H_{u}^{\x}}(t)\negthinspace\Bigl(r_{u}^{\x}(t)-\gamma_{u}^{\x}(t)\Bigr),
\]
\[
{Z_{u}^{\text{UL}}}^{2}\negthinspace(t+1)\negthinspace-\negthinspace{Z_{u}^{\text{UL}}}^{2}\negthinspace(t)\negthinspace\leq\negthinspace{p_{u}^{\text{UL}}}^{2}\negthinspace(t)\negthinspace+{\delta_{u}^{\text{UL}}}^{2}\negthinspace(t)\negthinspace-\negthinspace2{Z_{u}^{\text{UL}}}(t)\negthinspace\Bigl(\negthinspace\delta_{u}^{\text{UL}}\negthinspace(t)\negthinspace-\negthinspace p_{u}^{\text{UL}}(t)\negthinspace\Bigr),
\]
\[
{Z_{b}^{\text{DL}}}^{2}\negthinspace(t+1)\negthinspace-\negthinspace{Z_{b}^{\text{DL}}}^{2}\negthinspace(t)\negthinspace\leq\negthinspace{p_{b}^{\text{DL}}}^{2}\negthinspace(t)\negthinspace+{\delta_{b}^{\text{DL}}}^{2}\negthinspace(t)\negthinspace-\negthinspace2{Z_{b}^{\text{DL}}}(t)\negthinspace\Bigl(\negthinspace\delta_{b}^{\text{DL}}(t)\negthinspace-\negthinspace p_{b}^{\text{DL}}(t)\negthinspace\Bigr).
\]
After replacing the squared queue values in the drift-plus-penalty
function in (\ref{eq:drift_eq}) by the above inequalities, and given
the fact that the variables in the following term $\frac{1}{2}\mathbb{E}\{\sum_{u\in\mathcal{U}}\Bigl[{A_{u}^{\text{UL}}}^{2}+{A_{u}^{\text{DL}}}^{2}+{\delta_{u}^{\text{UL}}}^{2}+{p_{u}^{\text{UL}}}^{2}+{\gamma_{u}^{\text{UL}}}^{2}+{\gamma_{u}^{\text{DL}}}^{2}+2{r_{u}^{\text{UL}}}^{2}+2{r_{u}^{\text{DL}}}^{2}\Bigr]+\sum_{b\in\mathcal{B}}\Bigl[{\delta_{b}^{\text{DL}}}^{2}+{p_{b}^{\text{DL}}}^{2}\Bigr]\}$
are bounded above by constant values that are independent of $t$,
replacing this term by the upper bound constant $C$ yields equation
(\ref{eq:drift_plus_penalty_bound}).
\end{IEEEproof}

\subsection{Problem Decomposition Using the Lyapunov Framework}

Next, the problem in (\ref{eq:drift_plus_penalty_bound}) is solved
by choosing the control action that minimizes the terms in the right-hand
side at each time instant $t$. Since the optimization parameters
in each of the two terms in (\ref{eq:drift_plus_penalty_bound}) are
disjoint, each term can be decoupled and solved independently and
concurrently based on the observation of the system and virtual queues.

\subsubsection{Auxiliary variables selection}

The term $\#1$ in the drift-plus-penalty equation is related to the
auxiliary variable selection. By decoupling the problem per user,
the optimal values of the auxiliary variables are selected by solving
the following maximization problem:

\begin{subequations}\label{eq:aux_sub_prob}
\begin{alignat}{1}
\max_{\gamma_{u}^{\text{UL}},\gamma_{u}^{\text{DL}}} & v.f_{^{\text{UL}}}(\gamma_{u}^{\text{UL}}(t))-H_{u}^{\text{UL}}(t)\gamma_{u}^{\text{UL}}(t)\\
 & \quad\quad+v.f_{^{\text{DL}}}(\gamma_{u}^{\text{DL}}(t))-H_{u}^{\text{DL}}(t)\gamma_{u}^{\text{DL}}(t)\nonumber \\
\textrm{subject to } & \gamma_{u}^{\text{UL}}(t)\leq r_{\max}^{\text{UL}},\quad\gamma_{u}^{\text{DL}}(t)\leq r_{\max}^{\text{DL}}
\end{alignat}

\end{subequations}

We choose to maximize linear UL and DL utility functions, i.e., $f_{^{\text{UL}}}(\gamma_{u}^{\text{UL}})=(\gamma_{u}^{\text{UL}})$
and $f_{^{\text{DL}}}(\gamma_{u}^{\text{DL}})=(\gamma_{u}^{\text{DL}})$.
In this case, the parameter $v$ balances the tradeoff between maximizing
the instantaneous rate and maintaining fairness between users by prioritizing
users with higher queue sizes to ensure queue stability. Since (\ref{eq:aux_sub_prob})
is a linear function, the optimal value of the auxiliary variables
will equal their maximum value when $v-H_{u}^{\textrm{\x}}(t)$ is
not negative, and zero when it is negative, i.e.:

\begin{equation}
\gamma_{u}^{\x}(t)=\begin{cases}
r_{\max}^{\x} & H_{u}^{\textrm{\x}}(t)\leq v,\\
0 & \text{otherwise}.
\end{cases}
\end{equation}

\subsubsection{User scheduling and power allocation problem}

By substituting $r_{u}^{\text{UL}}$ and $r_{u}^{\text{DL}}$ from
(\ref{eq:UL_rate}) and (\ref{eq:DL_rate}), and $p_{b}^{\text{DL}}=\sum_{u\in\mathcal{U}}x_{bu}^{\text{DL}}p_{bu}^{\text{DL}}$,
the term $\#2$ in the drift-plus-penalty equation can be reformulated
as:

\begin{align}
 & \sum_{u\in\mathcal{U}}\biggr[\sum_{b\in\mathcal{B}}\Bigl[(Q_{u}^{\text{UL}}(t)+H_{u}^{\text{UL}}(t))x_{bu}^{\text{UL}}(t)R_{bu}^{\text{UL}}(t)\nonumber \\
 & \qquad\qquad+(Q_{u}^{\text{DL}}(t)+H_{u}^{\text{DL}}(t))x_{bu}^{\text{DL}}(t)R_{bu}^{\text{DL}}(t)\Bigr]\nonumber \\
 & \qquad+Z_{u}^{\text{UL}}(t)(\delta_{u}^{\text{UL}}(t)-p_{u}^{\text{UL}}(t))\biggl]\nonumber \\
 & +\sum_{b\in\mathcal{B}}\Bigl[Z_{b}^{\text{DL}}(t)(\delta_{b}^{\text{DL}}(t)-\sum_{u\in\mathcal{U}}x_{bu}^{\text{DL}}(t)p_{bu}^{\text{DL}}(t))\Bigr],\nonumber \\
 & =\sum_{u\in\mathcal{U}}\sum_{b\in\mathcal{B}}\Bigl[\Psi_{bu}^{\textrm{UL}}(t)+\Psi_{bu}^{\textrm{DL}}(t)\Bigr]+\Omega_{u}^{\textrm{UL}}(t)+\sum_{b\in\mathcal{B}}\Omega_{b}^{\textrm{DL}}(t).\label{eq:sched_and_power}
\end{align}

{} Therefore, the decomposed problem can be expressed as:

\begin{subequations}\label{eq:UAMS}

\begin{align}
\max_{X_{b}^{\text{UL}},X_{b}^{\text{DL}},\boldsymbol{p}^{\text{UL}},\{\boldsymbol{p}_{b}^{\text{DL}}\}} & \,\sum_{u\in\mathcal{U}}\biggr[\sum_{b\in\mathcal{B}}\Bigl[\Psi_{bu}^{\textrm{UL}}(t)+\Psi_{bu}^{\textrm{DL}}(t)\Bigr]\nonumber \\
 & \qquad+\Omega_{u}^{\textrm{UL}}(t)\biggl]+\sum_{b\in\mathcal{B}}\Omega_{b}^{\textrm{DL}}(t),\\
\textrm{ subject to} & \:(\ref{eq:const4})-(\ref{eq:const_feas}),
\end{align}

\end{subequations}

The above problem is a combinatorial problem of associating users
to SBSs and selecting whether to operate in HD-OMA, HD-NOMA, or FD-OMA
mode, and allocating UL and DL power, which has an exponential complexity.
To solve this problem, we decouple the user association and mode selection
problem from the power allocation problem. First, a low complexity
matching algorithm is proposed that matches users to SBSs and selects
the transmission mode. Given the outcome of the matching algorithm,
a power allocation problem is proposed to optimally allocate UL and
DL powers.

\section{User Association and Power Allocation \label{sec:rate_power_prob}}

In this section, the proposed solutions for the UAMS and UDPO problems
are presented. First, the UAMS problem objective is to solve the user
association and mode selection, which is formulated as a matching
game between users and SBSs assuming a fixed power allocation. Subsequently,
a local matching algorithm inspired from the Gale and Shapley algorithm
\cite{gale_shapley_college,gale_shapley} is proposed. To solve the
problem locally, knowledge of inter-cell interference is needed. Therefore,
a learning based inter-cell interference estimation method is devised.
Subsequently, the UDPO problem of allocating UL and DL powers is formulated
as a DC (difference of convex functions) programing problem and is
solved iteratively.

\subsection{User-SBS Matching}

\subsubsection{Matching preliminaries}

Matching theory is a framework that solves combinatorial problems
of matching members of two disjoint sets of players in which each
player is interested in matching to one or more player in the other
set \cite{gale_shapley,matching_mag}. Matching is performed on the
basis of preference profiles of players. Here, both SBSs and users
are assumed to have preferences towards each other to form \emph{matching
pair}s. Preferences of the sets of SBSs and users $(\mathcal{B},\mathcal{U}$),
denoted $\succ_{b}$ and $\succ_{u}$, represent each player ranking
of the players in the other set.
\begin{defn}
\label{def:matching}A many-to-one matching $\Upsilon$ is defined
as a mapping from the set $\mathcal{B}\cup\mathcal{U}$ into the set
of all subsets of $\mathcal{B}\cup\mathcal{U}$, where $\mathcal{B}=[1,\ldots,B]$
and $\mathcal{U}=[1,\ldots,U]$ are two disjoint sets of SBSs and
users, respectively, such that for each $b\in\mathcal{B}$ and $u\in\mathcal{U}$:
1)~$\Upsilon(b)\subseteq\mathcal{U}$; 2)~$\Upsilon(u)\in\mathcal{B}$;
3)~$\Upsilon(b)\leq q$; 4)~$\Upsilon(u)\leq1$; 5)~$\Upsilon(u)=\{b\}\Leftrightarrow u\in\Upsilon(b)$.
Note from 3) that an SBS is matched to at most a quota of $q$ users,
whereas from 4) that users are matched to at most one SBS.%
\end{defn}
\,
\begin{defn}
Preferences of Both SBSs and users are \emph{transitive}, such that
if $b\succ_{u}b'$ and $b'\succ_{u}b''$, then $b\succ_{u}b''$, and
similarly, if $u\succ_{b}u'$ and $u'\succ_{b}u''$, then $u\succ_{b}u''$.
\end{defn}
\,
\begin{defn}
\label{def:blocking_pair}Given a matching $\Upsilon$ and a pair
$(u,b)$ with $u\notin\Upsilon_{b}$ and $b\notin\Upsilon_{u}$, $(u,b)$
is said to be blocking the matching $\Upsilon$ and form a blocking
pair if: 1) $\mathcal{T}\succ_{b}\Upsilon_{b}$, $\mathcal{T}=\{u\}\cup\Upsilon_{b}$,
2) $b\succ_{u}b'$, $\Upsilon_{u}=b'$.
\end{defn}

\subsubsection{Preference profiles}

Matching allows defining preference profiles that capture the utility
function of the players. Accordingly, to construct preference profiles
that leads to maximizing the system-wide utility, SBSs and users rank
each other with the objective of maximizing the utility function in
(\ref{eq:UAMS}).%
{} Notice from (\ref{eq:UAMS}) that the utility function is a sum of
weighted rate maximization terms, i.e., ($\Psi_{bu}^{\x}$) and power
control terms, that are $\Omega_{u}^{\textrm{UL}}(t)$ and $\Omega_{b}^{\textrm{DL}}(t)$.
Essentially, the preference of user $u$ should reflect the user preference
to maximize its own UL and DL rate, which is expressed as:

\begin{equation}
b\succ_{u}b'\Leftrightarrow\sum_{l\in\{\textrm{UL},\textrm{DL}\}}\Psi_{bu}^{\x}>\sum_{l\in\{\textrm{UL},\textrm{DL}\}}\Psi_{b'u}^{\x}.\label{eq:user_preference}
\end{equation}

On the other hand, each SBS aims to maximize the rate of its serving
users as well as maintaining the UL and DL power queues stable through
optimizing its user and mode selection. Therefore, SBSs rank the subsets
of proposing users (including the individual users). The proposed
preference of SBS $b$ over two subsets of users $\mathcal{M\subseteq\mathcal{U}}$
and $\mathcal{M}'\subseteq\mathcal{U}$ is defined as:

\allowdisplaybreaks[0]
\begin{multline}
\mathcal{M}\succ_{b}\mathcal{M}'\\
\Leftrightarrow\Omega_{b}^{\textrm{DL}}(\mathcal{M})+\sum_{\substack{u\in\mathcal{M}\\
l\in\{\textrm{UL},\textrm{DL}\}
}
}\Bigl[\Psi_{bu}^{\x}(\mathcal{M})+\Omega_{u}^{\x}(\mathcal{M})\Bigr]>\\
\Omega_{b}^{\textrm{DL}}(\mathcal{M}')+\sum_{\substack{u\in\mathcal{M}'\\
l\in\{\textrm{UL},\textrm{DL}\}
}
}\Bigl[\Psi_{bu}^{\x}(\mathcal{M}')+\Omega_{u}^{\x}(\mathcal{M}')\Bigr].
\end{multline}
\allowdisplaybreaks

\subsubsection{Matching algorithm}

A matching algorithm to solve the many-to-one matching problem is
introduced. The algorithm is a modified version of the \emph{deferred
acceptance (DA)} matching algorithm \cite{gale_shapley}. In each
round of the proposed algorithm, each unmatched user proposes to the
top SBS in its preference list, SBSs then accept or reject the offers.
Due to the fact that an SBS accepts only users who maximize its own
utility and rejects others, it eliminates the possibility of having
blocking pairs, since it would have formed blocking pairs with these
users otherwise. The detailed matching algorithm is presented in Algorithm
\ref{alg:matching}.

\begin{algorithm}[tbh]
\begin{algorithmic}[1]\footnotesize

\STATE  \textbf{Initialization: }all users start unmatched, SBSs
initialize empty lists of proposals.

\STATE  Users construct their own preference lists following (\ref{eq:user_preference}).

\STATE  \textbf{repeat} each unmatched user with nonempty preference
list propose to its most preferred SBS.

\STATE    \textbf{foreach} SBS $j$ \textbf{do}.

 \begin{ALC@g} \STATE  SBS observes the subset of users that are
in its proposal list, denoted as $\mathcal{M}_{j}$.

\STATE  \textbf{if} $\left|\mathcal{M}_{j}\right|=1$,

\begin{ALC@g}\STATE The user in $\mathcal{M}_{j}$ is accepted.

\end{ALC@g}\STATE  \textbf{elseif} $\left|\mathcal{M}_{j}\right|=2$,

\begin{ALC@g}\STATE   Both users in the list are accepted only if
$\mathcal{M}_{j}\succ_{b}\mathcal{M}_{j}',\forall\mathcal{M}_{j}'\subseteq\mathcal{M}_{j}$.

\STATE   Otherwise, the user in the most preferred $\mathcal{M}_{j}'$
is accepted and the other is rejected.

\end{ALC@g}\STATE  \textbf{elseif }$\left|\mathcal{M}_{j}\right|>2$,

  \begin{ALC@g}\STATE  SBS identifies the feasible subsets $\mathcal{M}_{j}'\subseteq\mathcal{M}_{j}$
that satisfy constraints (\ref{eq:const7}), (\ref{eq:const8}), and
(\ref{eq:const_feas}).

\STATE  SBS calculates its preference of all the feasible subsets.

\STATE SBS accepts the users in the highest preferred subset, other
users are rejected.

 \end{ALC@g}\STATE  \textbf{end if}

\STATE  An accepted user is marked as matched.

\STATE  A rejected user removes SBS $j$ from its preference list.

 \end{ALC@g}\STATE  end \textbf{foreach}

 \STATE  \textbf{until }all users are either matched or not having
SBSs remaining in their preference lists.\textbf{ }

\STATE  \textbf{Output: }a stable matching $\Upsilon$.

\end{algorithmic}

\caption{\label{alg:matching}User-SBS Matching Algorithm (a modified DA algorithm
\cite{gale_shapley}) }
\end{algorithm}

\begin{rem}
For a fixed preference profile, Algorithm \ref{alg:matching} is guaranteed
to find a two-sides stable matching between SBSs and users \cite{gale_shapley_college}.
\end{rem}
The above remark states that if the preference profiles of some of
the players are not affected by the current matching of other players,
the DA-based matching algorithm will converge to a stable matching.
However, since the preference profiles are function of the instantaneous
service rates, there exist interdependencies between the player's
preferences, which is known as matching \emph{externalities}. With
these externalities, DA-based algorithms are not guaranteed to find
a stable matching. Moreover, the dependency of the preference on other
players' current matching makes it necessary for SBSs to communicate
with each other to calculate the system-wide utility. The overhead
due to this communication grows as the number of users and SBSs increase,
rendering the matching algorithm impractical. To this end, a utility
estimation procedure through inter-cell interference learning is carried
out in the following subsection.

\subsubsection{Dealing with externalities}

Having the instantaneous rate in the term $\Psi_{bu}$ causes the
preferences of players to vary as the matching changes, due to having
both the intra-cell and inter-cell interference as per equations (\ref{eq:UL_rate})-(\ref{eq:SINR_DL}).
Consequently, externalities are introduced to the matching problem,
which makes it a computationally hard task to find a stable matching.
Instead, we propose an inter-cell interference learning method that
allows SBSs and users to locally estimate their utility at each time
instant $t$. Hence, externalities due to varying inter-cell interference
are avoided.

Under this procedure, all SBSs and users keep record of the inter-cell
interference experienced at each time instant. Let the measured inter-cell
interference at SBS $b$ and user $u$ in the time instant $t-1$
be $\tilde{I}_{b,\textrm{inter}}(t-1)$ and $\tilde{J}_{u,\textrm{inter}}(t-1)$,
respectively, and $\hat{I}_{b,\textrm{inter}}(t)$ and $\hat{J}_{u,\textrm{inter}}(t)$
are the estimated inter-cell interference at time instant $t$. A
time-average estimation of the inter-cell interference is performed
as follows:

\begin{alignat}{1}
\hat{I}_{b,\textrm{inter}}(t) & \negthinspace=\negthinspace\nu_{1}(t)\tilde{I}_{b,\textrm{inter}}(t\negthinspace-\negthinspace1)\negthinspace+\negthinspace(1\negthinspace-\negthinspace\nu_{1}(t))\hat{I}_{b,\textrm{inter}}(t\negthinspace-\negthinspace1),\\
\hat{J}_{u,\textrm{inter}}(t) & \negthinspace=\negthinspace\nu_{2}(t)\tilde{J}_{u,\textrm{inter}}(t\negthinspace-\negthinspace1)\negthinspace+\negthinspace(1\negthinspace-\negthinspace\nu_{2}(t))\hat{J}_{u,\textrm{inter}}(t\negthinspace-\negthinspace1),
\end{alignat}
where $\nu_{1}$and $\nu_{2}$ are learning parameters.

Once the learning procedure has been established, the preference profiles
of players are built using the estimated terms. First, we assume that
players build their preference profiles over SBSs using the estimated
inter-cell interference to calculate the estimated weighted rate term,
denoted as $\hat{\Psi}_{bu}^{\x}$. This assumption avoids the complexity
of taking into account the intra-cell interference in calculating
users' utility, in which the preference will vary significantly according
to the SBS current mode selection as well as the SIC order if NOMA
is selected. Hence, the user preference can be expressed as:

\begin{equation}
b\succ_{u}b'\Leftrightarrow\sum_{l\in\{\textrm{UL},\textrm{DL}\}}\hat{\Psi}_{bu}^{\x}>\sum_{l\in\{\textrm{UL},\textrm{DL}\}}\hat{\Psi}_{b'u}^{\x}.
\end{equation}
On the other hand, as a set of more than one user can propose to an
SBS, the preference of a subset of users over another has to take
into account the actual intra-cell interference, as well as the estimated
inter-cell interference. Therefore, the estimated UL and DL SINR values
of user $u$ at SBS \textbf{$b$} are expressed as:

\begin{align}
\textrm{\ensuremath{\hat{\Gamma}}}_{bu}^{\textrm{UL}}(t) & =\frac{p_{u}^{\text{UL}}(t)h_{bu}(t)}{N_{0}+\hat{I}_{b,\textrm{inter}}(t)+I_{b,\textrm{intra}}(t)+p_{b}^{\text{DL}}(t)\negthinspace/\negthinspace\zeta},\label{eq:appr_SINR_UL}\\
\textrm{\ensuremath{\hat{\Gamma}}}_{bu}^{\textrm{DL}}(t) & =\frac{p_{bu}^{\text{DL}}(t)h_{bu}(t)}{N_{0}+\hat{J}_{u,\textrm{inter}}(t)+J_{u,\textrm{intra}}(t)},\label{eq:appr_SINR_DL}
\end{align}
where $I_{b,\textrm{intra}}$ and $J_{u,\textrm{intra}}$ are the
actual intra-cell interference resulting from FD or NOMA operation.
Subsequently, the preference of SBS $b$ over two subsets of users
$\mathcal{M}$ and $\mathcal{M}'$ is defined as:
\begin{multline}
\mathcal{M}\succ_{b}\mathcal{M}'\\
\Leftrightarrow\Omega_{b}^{\textrm{DL}}(\mathcal{M})+\sum_{\substack{u\in\mathcal{M}\\
l\in\{\textrm{UL},\textrm{DL}\}
}
}\Bigl[\hat{\Psi}_{bu}^{\x}(\mathcal{M})+\Omega_{u}^{\x}(\mathcal{M})\Bigr]>\\
\Omega_{b}^{\textrm{DL}}(\mathcal{M}')+\sum_{\substack{u\in\mathcal{M}'\\
l\in\{\textrm{UL},\textrm{DL}\}
}
}\Bigl[\hat{\Psi}_{bu}^{\x}(\mathcal{M}')+\Omega_{u}^{\x}(\mathcal{M}')\Bigr].
\end{multline}

\subsubsection{Analysis of matching stability}
\begin{defn}
Matching $\Upsilon$ is pairwise stable if it is not blocked by any
pair that does not exist in $\Upsilon$.
\end{defn}
\begin{lem}
Algorithm \ref{alg:matching} converges to a pairwise-stable matching
$\Upsilon^{*}$.
\end{lem}
\begin{IEEEproof}
Suppose that there exists an arbitrary pair $(u,b)$ which does not
exist in $\Upsilon^{*}$ in which $\mathcal{T}\succ_{b}\Upsilon_{b}^{*}$,
$\mathcal{T}=\{u\}\cup\Upsilon_{b}^{*}$, and $b\succ_{u}b'$, $\Upsilon_{u}^{*}=b'$.
Since $b\succ_{u}b'$, this implies that user $u$ has proposed to
$b$ and got rejected at an intermediate round $i$ before being matched
to $b'$. Accordingly, $\Upsilon_{b}^{i}\succ_{b}\{u\}\cup\Upsilon_{b}^{i}$.
As~$\Upsilon_{b}^{*}\succ_{b}\Upsilon_{b}^{i}$, this implies that
$u\notin\Upsilon_{b}^{*}$ since the preference lists are transitive,
which contradicts our supposition. Therefore, the matching $\Upsilon^{*}$
is not blocked by any pair, i.e., is pairwise-stable.
\end{IEEEproof}
\begin{thm}
Algorithm \ref{alg:matching} converges to a pairwise-stable matching
$\Upsilon^{*}$ in a finite number of rounds.
\end{thm}
\begin{IEEEproof}
Since each user can propose to an SBS only once, removing it from
its preference list if it is rejected, it is easy to see that a user
$u$ has a preference list of maximum length of \textbf{$B$}. As
more rounds are conducted, the preference lists grow smaller. This
means that each user performs a maximum of $B$ proposals. After which
the algorithm will converge to a pairwise-stable matching $\Upsilon^{*}$.
\end{IEEEproof}

\subsection{UL/DL Power Optimization Problem (UDPO)}

After users are associated to SBSs and transmission modes are selected,
power allocation is performed to optimize the UL and DL power levels
for each scheduled UE and SBS. We assume that a central-controller
performs the optimal power allocation. The central-controller requires
channel knowledge of only the scheduled users. This substantially
reduces the complexity of the channel reporting, as compared to a
fully centralized solution. Given the user association obtained from
the matching algorithm, equation (\ref{eq:sched_and_power}) can be
rewritten as follows:

\begin{align}
 & \sum_{b\in\mathcal{B}}\sum_{u\in\Upsilon_{b}(t)}\Bigl[\Psi_{bu}^{\textrm{UL}}(t)+\Psi_{bu}^{\textrm{DL}}(t)+\Omega_{u}^{\textrm{UL}}(t)\Bigr]+\sum_{b\in\mathcal{B}}\Omega_{b}^{\textrm{DL}}(t)\nonumber \\
 & =\sum_{b\in\mathcal{B}}\biggr[\sum_{u\in\Upsilon_{b}(t)}\Bigl[w_{u}^{\text{UL}}(t)R_{bu}^{\text{UL}}(t)+w_{u}^{\text{DL}}(t)R_{bu}^{\text{DL}}(t)\nonumber \\
 & \qquad\qquad\qquad\quad+Z_{u}^{\text{UL}}(t)(\delta_{u}^{\text{UL}}(t)-p_{u}^{\text{UL}}(t))\Bigr]\nonumber \\
 & \quad\quad\quad+Z_{b}^{\text{DL}}(t)(\delta_{b}^{\text{DL}}(t)-\sum_{u\in\Upsilon_{b}(t)}p_{bu}^{\text{DL}}(t))\biggl],\label{eq:power_opt}
\end{align}
where $w_{u}^{\x}=Q_{u}^{\x}+H_{u}^{\text{\x}}$. The power optimization
problem is expressed as\footnote{For the sake of brevity, we omit the time index $t$ from this power
optimization subproblem, as it is performed each time instant. }:

\begin{subequations}

\begin{align}
\max_{\boldsymbol{p}=\{\boldsymbol{p}^{\text{UL}},\{\boldsymbol{p}_{b}^{\text{DL}}\}\}} & \sum_{b\in\mathcal{B}}\sum_{u\in\Upsilon_{b}}\negthinspace\Bigl[\Psi_{bu}^{\textrm{UL}}(\boldsymbol{p})\negthinspace+\negthinspace\Psi_{bu}^{\textrm{DL}}(\boldsymbol{p})\negthinspace+\negthinspace\Omega_{u}^{\textrm{UL}}(\boldsymbol{p})\Bigr]\negthinspace+\negthinspace\Omega_{b}^{\textrm{DL}}(\boldsymbol{p}),\\
 & p_{u}^{\text{UL}}\leq P_{\max}^{\text{UL}},\;\forall u\in\Upsilon,\\
 & p_{b}^{\text{DL}}\leq P_{\max}^{\text{DL}},\;\forall b\in\Upsilon.\\
 & Y_{uu'}\geq0\;\;\forall u,u'\in\Upsilon,
\end{align}

\end{subequations}

The above problem is a non-concave utility maximization problem, due
to the interference terms in the UL and DL rate expressions. This
makes it complex to solve, even with a central controller, with complexity
that increases with the number of SBSs and users. To obtain an efficient
solution, we transform the problem into a DC programming optimization
problem and solve it iteratively \cite{CCCP_ref}. First, we substitute
the service rate terms in the objective function:\vspace{-0.2cm}
\begin{align}
 & =\sum_{b\in\mathcal{B}}\biggr[\sum_{u\in\Upsilon_{b}(t)}\Bigl[w_{u}^{\text{UL}}f_{b}\log_{2}(1\!+\!\textrm{\ensuremath{\Gamma}}_{bu}^{\textrm{UL}}(\boldsymbol{p}))\nonumber \\
 & \qquad\qquad\qquad\;+\!w_{u}^{\text{DL}}f\log_{2}(1\!+\!\textrm{\ensuremath{\Gamma}}_{bu}^{\textrm{DL}}(\boldsymbol{p}))\nonumber \\
 & \qquad\qquad\qquad\;+\Omega_{bu}^{\textrm{UL}}(\boldsymbol{p})\,\Bigr]+\Omega_{b}^{\textrm{DL}}(\boldsymbol{p})\,\biggl],\label{eq:UDPO}
\end{align}
Then, the SINR expressions from (\ref{eq:SINR_UL}) and (\ref{eq:SINR_DL})
is rewritten as:

\begin{align}
\textrm{\ensuremath{\Gamma}}_{bu}^{\textrm{UL}}(\boldsymbol{p}) & =\frac{p_{u}^{\text{UL}}h_{bu}}{N_{0}+I_{bu,\textrm{UL}}(\boldsymbol{p})},\label{eq:SINR_UL-2}\\
\textrm{\ensuremath{\Gamma}}_{bu}^{\textrm{DL}}(\boldsymbol{p}) & =\frac{p_{bu}^{\text{DL}}h_{bu}}{N_{0}+I_{bu,\textrm{DL}}(\boldsymbol{p})},\label{eq:SINR_DL-2}
\end{align}
where $I_{bu,\textrm{UL}}=I_{b}^{\textrm{UL-UL}}+I_{b}^{\textrm{DL-UL}}+I_{bu}^{\textrm{NOMA-UL}}+p_{b}^{\text{DL}}/\zeta$
and $I_{bu,\textrm{DL}}=I_{u}^{\textrm{DL-DL}}+I_{u}^{\textrm{UL-DL}}+I_{bu}^{\textrm{NOMA-DL}}$.

By substituting (\ref{eq:SINR_UL-2}) and (\ref{eq:SINR_DL-2}) in
(\ref{eq:UDPO}), the objective function becomes:\vspace{-0.3cm}

\begin{align}
 & =\sum_{b\in\mathcal{B}}\biggr[\sum_{u\in\Upsilon_{b}(t)}\Bigl[w_{u}^{\text{UL}}f_{b}\Bigl(\log_{2}(N_{0}+p_{u}^{\text{UL}}h_{bu}+I_{bu,\textrm{UL}}(\boldsymbol{p}))\nonumber \\
 & \qquad\qquad\qquad-\log_{2}(N_{0}+I_{bu,\textrm{UL}}(\boldsymbol{p}))\Bigr)\nonumber \\
 & \qquad\qquad\qquad+w_{u}^{\text{DL}}f_{b}\Bigl(\log_{2}(N_{0}+p_{b}^{\text{DL}}h_{bu}+I_{bu,\textrm{DL}}(\boldsymbol{p}))\nonumber \\
 & \qquad\qquad\qquad-\log_{2}(N_{0}+I_{bu,\textrm{DL}}(\boldsymbol{p}))\Bigr)\nonumber \\
 & \qquad\qquad\qquad+\Omega_{u}^{\textrm{UL}}(\boldsymbol{p})\Bigr]+\Omega_{b}^{\textrm{DL}}(\boldsymbol{p})\biggl],
\end{align}
\begin{align}
 & =\sum_{b\in\mathcal{B}}\biggr[\sum_{u\in\Upsilon_{b}}\Bigl[\stackrel{\textrm{concave}}{\overbrace{\text{F}_{u}^{\text{UL}}(\boldsymbol{p})}}+\stackrel{\textrm{convex}}{\overbrace{\text{G}_{u}^{\text{UL}}(\boldsymbol{p})}}+\stackrel{\textrm{concave}}{\overbrace{\text{F}_{u}^{\text{DL}}(\boldsymbol{p})}}+\stackrel{\textrm{convex}}{\overbrace{\text{G}_{u}^{\text{DL}}(\boldsymbol{p})}}\nonumber \\
 & \qquad\qquad\qquad+\stackrel{\textrm{affine}}{\overbrace{\Omega_{u}^{\textrm{UL}}(\boldsymbol{p})}}\Bigr]+\stackrel{\textrm{affine}}{\overbrace{\Omega_{b}^{\textrm{DL}}(\boldsymbol{p})}}\biggl].
\end{align}

Finally, we use the convex-concave procedure \cite{CCCP_boyd} to
convexify the above problem. The convex functions are first substituted
by their first order linear approximation at iteration~$i_{t}$:

\begin{equation}
\tilde{\text{G}_{u}^{\text{\x}}}(\boldsymbol{p},\boldsymbol{p}(i_{t}))=G_{u}^{\x}(\boldsymbol{p},\boldsymbol{p}(i_{t}))+\nabla G_{u}^{\x}(\boldsymbol{p},\boldsymbol{p}(i_{t}))^{T}\left(\boldsymbol{p}-\boldsymbol{p}(i_{t})\right).\label{eq:ccp}
\end{equation}

\begin{algorithm}[tbh]
\begin{algorithmic}[1]\footnotesize

\STATE  \textbf{Initialization: }find an initial feasible point,
set $i_{t}=0$.

\STATE  \textbf{repeat}

\STATE  calculate $\tilde{\text{G}_{u}^{\text{\x}}}(\boldsymbol{p},\boldsymbol{p}(i_{t})),\forall u\in\Upsilon_{b}$
from (\ref{eq:ccp}).

\STATE solve the convex minimization problem using the interior point
method.

\STATE  update iteration: $i_{t}=i_{t}+1$.

\STATE  \textbf{until }the utility improvement $\leq\delta$.\textbf{ }

\STATE  \textbf{Output: }optimal power vectors: $\boldsymbol{p}=\{\boldsymbol{p}^{\text{UL}},\{\boldsymbol{p}_{b}^{\text{DL}}\}\}$.

\end{algorithmic}

\caption{\label{alg:power_opt}Iterative Power Allocation Algorithm (UDPO)
at time instant $t$}
\end{algorithm}

\begin{figure}[htbp]
\includegraphics[width=1\columnwidth]{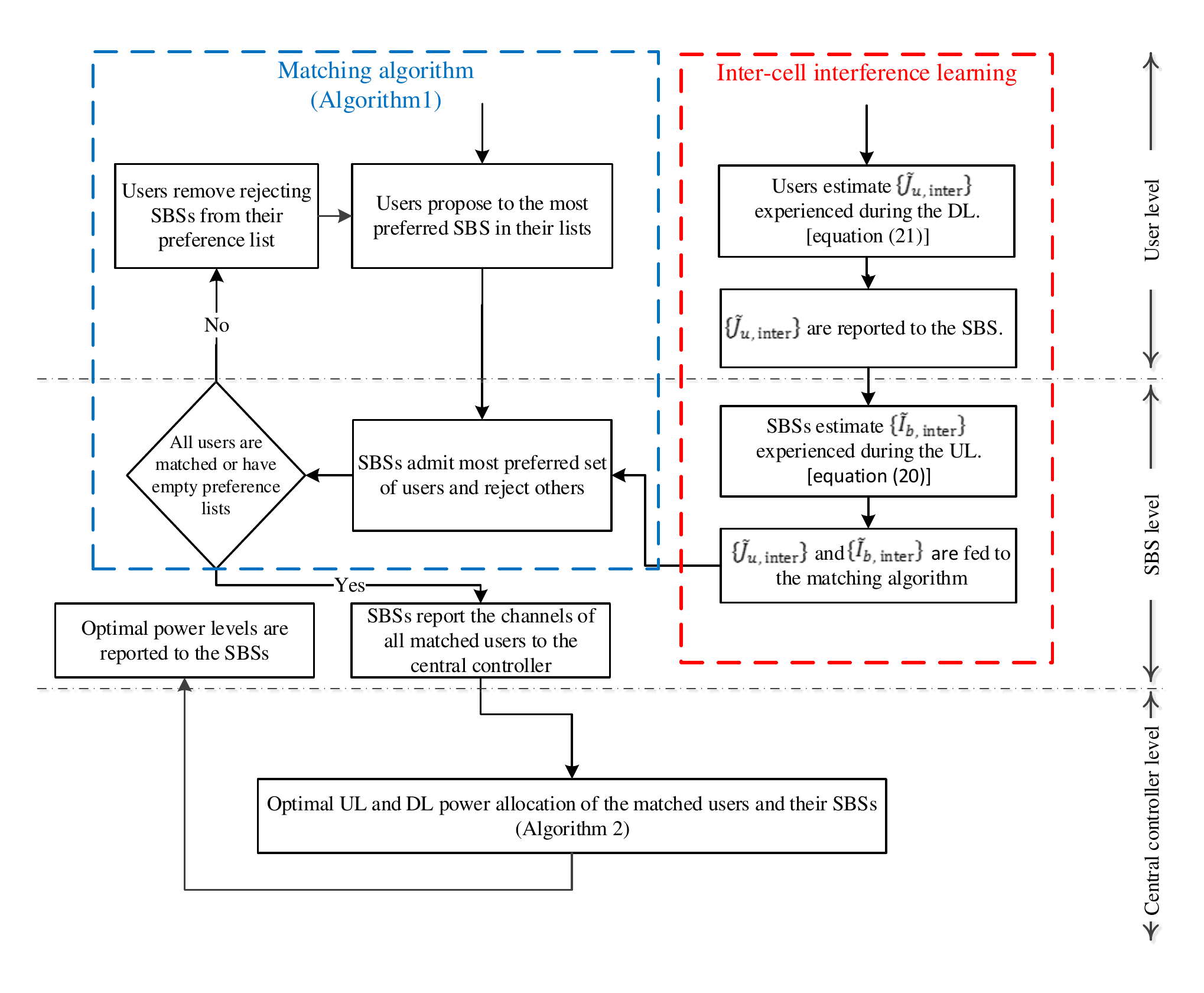}
\centering{}\caption{\label{fig:FD_NOMA_flowchart}A flowchart illustrating the sequence,
implementation and connections between the proposed learning scheme,
matching algorithm and power allocation scheme.}
\end{figure}

The problem is then solved iteratively, as depicted in Algorithm 2.
The feasibility problem \cite{boyd_book} is solved to find a feasible
initial starting point. In this problem, the utility function is set
to zero such that if the optimal value is zero, the solution fits
as an initial point that satisfies the problem constraints. In the
subsequent iterations, the initial point is always feasible, as it
is the optimal solution of the previous iteration.
\begin{thm}
Algorithm \ref{alg:power_opt} generates a sequence of $\boldsymbol{p}(i_{t})$
that converges to a local optimum $\boldsymbol{p}^{*}$.
\end{thm}
\begin{IEEEproof}
Let the utility function $f_{p}(\boldsymbol{p})=\stackrel{\textrm{concave}}{\overbrace{\text{F}_{u}^{\x}(\boldsymbol{p})}}+\stackrel{\textrm{convex}}{\overbrace{\text{G}_{u}^{\text{\x}}(\boldsymbol{p})}}+\stackrel{\textrm{affine}}{\overbrace{\Omega_{u}^{\textrm{\x}}(\boldsymbol{p})}}$
be the function to be maximized. At iteration $i_{t}$, the convex
function $\text{G}_{u}^{\text{\x}}(\boldsymbol{p},)$ is replaced
by the affine function $\tilde{\text{G}_{u}^{\text{\x}}}(\boldsymbol{p},\boldsymbol{p}(i_{t}))$
to make the objective function concave with respect to the reference
point $\boldsymbol{p}(i_{t})$. Accordingly, we have a concave utility
function at iteration $i_{t}$, that can be solved efficiently and
results in the solution $\boldsymbol{p}^{*}(i_{t})$. At this point,
the original utility function can be expressed as $f_{p}(\boldsymbol{p}^{*}(i_{t}))=\text{F}_{u}^{\x}(\boldsymbol{p}^{*}(i_{t}))+\text{G}_{u}^{\text{\x}}(\boldsymbol{p}^{*}(i_{t}))+\Omega_{u}^{\textrm{\x}}(\boldsymbol{p}^{*}(i_{t}))$.
Subsequently, the obtained solution is utilized at iteration $i_{t}+1$
such that $\tilde{\text{G}_{u}^{\text{\x}}}(\boldsymbol{p},\boldsymbol{p}(i_{t}+1))=\tilde{\text{G}_{u}^{\text{\x}}}(\boldsymbol{p},\boldsymbol{p}^{*}(i_{t}))$,
in which the optimal solution will be $\boldsymbol{p}^{*}(i_{t}+1)$
and the corresponding utility of the original problem is $f_{p}(\boldsymbol{p}^{*}(i_{t}+1))=\text{F}_{u}^{\x}(\boldsymbol{p}^{*}(i_{t}+1))+\text{G}_{u}^{\text{\x}}(\boldsymbol{p}^{*}(i_{t}+1))+\Omega_{u}^{\textrm{\x}}(\boldsymbol{p}^{*}(i_{t}+1))$.

By comparing the original utility functions in the two subsequent
iterations, we can find that:\vspace{-0.5cm}

\begin{align*}
\negthinspace\negthinspace f_{p}(\boldsymbol{p}\negthinspace^{*}(i_{t})\negthinspace)\negthinspace & \negthinspace=\negthinspace\text{F}_{u}^{\x}(\boldsymbol{p}\negthinspace^{*}(i_{t}))\negthinspace+\negthinspace\text{G}_{u}^{\text{\x}}(\boldsymbol{p}\negthinspace^{*}(i_{t}))\negthinspace+\negthinspace\Omega_{u}^{\textrm{\x}}(\boldsymbol{p}\negthinspace^{*}(i_{t}))\\
 & \negthinspace=\negthinspace\text{F}_{u}^{\x}(\boldsymbol{p}\negthinspace^{*}(i_{t}))\negthinspace+\negthinspace\tilde{\text{G}_{u}^{\text{\x}}}(\boldsymbol{p}\negthinspace^{*}(i_{t}),\boldsymbol{p}(i_{t}+1))\negthinspace+\negthinspace\Omega_{u}^{\textrm{\x}}(\boldsymbol{p}\negthinspace^{*}(i_{t}))\\
 & \negthinspace\leq\negthinspace\text{F}_{u}^{\x}(\boldsymbol{p}\negthinspace^{*}(\negthinspace i_{t}\negthinspace+\negthinspace1\negthinspace))\negthinspace+\negthinspace\tilde{\text{G}_{u}^{\text{\x}}}(\boldsymbol{p}\negthinspace^{*}(\negthinspace i_{t}\negthinspace+\negthinspace1\negthinspace),\negthinspace\boldsymbol{p}(\negthinspace i_{t}\negthinspace+\negthinspace1\negthinspace))\negthinspace+\negthinspace\Omega_{u}^{\textrm{\x}}\negthinspace(\boldsymbol{p}\negthinspace^{*}\negthinspace(\negthinspace i_{t}\negthinspace+\negthinspace1\negthinspace))\\
 & \negthinspace\leq\negthinspace\text{F}_{u}^{\x}(\boldsymbol{p}\negthinspace^{*}(i_{t}+1))\negthinspace+\negthinspace\text{G}_{u}^{\text{\x}}(\boldsymbol{p}\negthinspace^{*}(i_{t}+1))\negthinspace+\negthinspace\Omega_{u}^{\textrm{\x}}(\boldsymbol{p}\negthinspace^{*}(i_{t}+1))\\
 & \negthinspace=\negthinspace f_{p}(\boldsymbol{p}\negthinspace^{*}(i_{t}+1)),
\end{align*}
where the first inequality is due to the lower term being the optimal
(maximum) solution to the problem at iteration $i_{t}+1$, and the
second inequality is due to the convexity of $\tilde{\text{G}_{u}^{\text{\x}}}(\boldsymbol{p},\boldsymbol{p}(i_{t}+1))\leq\text{G}_{u}^{\text{\x}}(\boldsymbol{p})$.
Therefore, Algorithm \ref{alg:power_opt} will generate a sequence
of $\boldsymbol{p}(i_{t})$ that leads to a non-decreasing utility
function, i.e., $f_{p}(\boldsymbol{p}^{*}(1))\leq\cdots\leq f_{p}(\boldsymbol{p}^{*}(i_{t}))\leq f_{p}(\boldsymbol{p}^{*}(i_{t}+1))$.
Finally, due to the bounded constraints, the utility function is bounded,
and will converge to a solution that is local optimal.
\end{IEEEproof}
A flowchart of the UAMS and the UDPO problems is presented in Figure
\ref{fig:FD_NOMA_flowchart}. Next, we analyze the complexity of the
proposed scheme in terms of signaling overhead.

\subsection{Complexity Analysis}

First, to analyze the complexity of Algorithm \ref{alg:matching},
we investigate the maximum number of request signals coming from users'
proposal to an arbitrary SBS in a time instant. Let $\mathcal{U}_{b}\subseteq\mathcal{U}$
be the set of users under the coverage of SBS $b$. In the worst case
scenario, let all the users have SBS \textbf{$b$} as their most preferred
SBS. Hence, the SBS will receive requests from all the users in the
first iterations, and it has to accept at most a single user in HD-OMA
operation, a pair of users in FD-OMA mode, or a maximum of $q$ users
in HD-NOMA mode. The worst case will occur if, at each iteration,
the SBS is only accepting one user for HD-OMA operation and rejecting
the others. In this case, the maximum number of iterations will be
$|\mathcal{U}_{b}|$, and the total number of proposals is $|\mathcal{U}_{b}|+(|\mathcal{U}_{b}|-1)+\cdots+1=|\mathcal{U}_{b}|(|\mathcal{U}_{b}|+1)/2$.
Therefore, the complexity of Algorithm \ref{alg:matching} is $\mathcal{O}(|\mathcal{U}_{b}|^{2}).$

Next we analyze the complexity of Algorithm \ref{alg:power_opt}.
Since the power allocation is performed only over the subset of users
that are selected using Algorithm \ref{alg:matching}, the maximum
number of users from an SBS will be in the HD-NOMA mode, which is
equivalent to the quota $q$. Then, the maximum number of users in
the network (assuming all SBSs are in operation) will be $q^{\textrm{tot}}=|\mathcal{B}|q$.
Then, an SBS needs to report the channel between its $q$ users and
the other $q^{\textrm{tot}}-q$ users. The total amount of signaling
will be $=|\mathcal{B}|q(q^{\textrm{tot}}-q)=|\mathcal{B}|q(|\mathcal{B}|q-q)=|\mathcal{B}|q^{2}(|\mathcal{B}|-1)$
which results in a complexity of $\mathcal{O}\bigl((|\mathcal{B}|q)^{2}\bigr).$

\section{Simulation Results\label{sec:sim_res}}

In this section, we present simulation results and analysis to evaluate
the performance of the proposed framework. For benchmarking, we consider
the following baseline schemes:
\begin{enumerate}
\item \emph{HD-OMA scheme}: users are associated to the nearest SBS and
are allocated orthogonal resources for UL and DL. Requests are served
using a round robin (RR) scheduler.
\item \emph{HD-NOMA only scheme}: users are associated to the nearest SBS
and RR scheduler is used to serve UL and DL queues. If there are multiple
users in a scheduling queue, they are ranked according to their channel
gains, and are served using NOMA if the ratio between their channel
gains is at least $2$, otherwise, OMA is used. Power is allocated
to NOMA users based on their channel ranking, in uniform descending
order for UL NOMA and uniform ascending order for DL NOMA. SBSs operate
either in UL or in DL depending on the queue length on each link direction.
\item \emph{FD-OMA scheme}: users are associated to the nearest SBS and
a pair of users is served in FD mode if the channel gain between them
is greater than a certain threshold, otherwise users are served in
HD mode using RR scheduler.
\item \emph{Uncoordinated scheme}: in this scheme, users can be served in
either HD or FD and in OMA or NOMA modes. The proposed matching algorithm,
with inter-cell interference learning, is used for mode selection
and user association. Power is assumed to be fixed for OMA and is
similar to that of baseline $2$ for NOMA. This baseline represents
an uncoordinated version of the proposed scheme.
\end{enumerate}

\subsection{Simulation setup}

We consider an outdoor cellular network where SBSs are distributed
uniformly over the network area and users are distributed uniformly
within small cells area. An SBS is located at the center of each small
cell. SBS operate in TDD HD or in FD. If HD is used, they can operate
in NOMA or OMA to serve multiple users. A packet-level simulator is
considered and each experiment is run for $4000$ time subframes,
which are sufficient for the traffic and virtual queues to get stable.
Each result is averaged over $30$ different network topologies. The
main simulation parameters are presented in Table \ref{tab:sim_par}.

\begin{table}[htbp]
\caption{Simulation parameters\label{tab:sim_par}}

\centering{}%
\begin{tabular}{|l|l|}
\hline
\textbf{Parameter} & \textbf{Value}\tabularnewline
\hline
\hline
System bandwidth & 10 MHz\tabularnewline
\hline
Duplex modes & TDD HD/ FD\tabularnewline
\hline
Multiplexing mode & OMA/NOMA\tabularnewline
\hline
Sub-frame duration & 1 ms\tabularnewline
\hline
Network size & $500\times500\:m^{2}$ \tabularnewline
\hline
Small cell radius & $40m$\tabularnewline
\hline
Max. SBS transmit power & $22$ dBm\tabularnewline
\hline
Max. user transmit power  & $20$ dBm\tabularnewline
\hline
Path loss model & Multi-cell pico scenario \cite{standard_tdd}\tabularnewline
\hline
Shadowing standard deviation & $4$ dB\tabularnewline
\hline
Penetration loss & $0$ dB\tabularnewline
\hline
SI cancellation capability & $110$ dB\tabularnewline
\hline
Max. quota of NOMA users $q$ & $5$\tabularnewline
\hline
\multicolumn{2}{|c|}{\textbf{Lyapunov parameters}}\tabularnewline
\hline
$v$ parameter & $5\times10^{7}$\tabularnewline
\hline
UL power threshold $\delta_{u}^{\text{UL}}$ & $0.5\times P_{\max}^{\text{UL}}$\tabularnewline
\hline
DL power threshold $\delta_{b}^{\text{DL}}$ & $0.9\times P_{\max}^{\text{DL}}$\tabularnewline
\hline
\multicolumn{2}{|c|}{\textbf{Learning parameters}}\tabularnewline
\hline
SBS learning parameter $\nu_{1}$ & $0.1$\tabularnewline
\hline
User learning parameter $\nu_{2}$ & $0.1$\tabularnewline
\hline
\end{tabular}
\end{table}

\subsection{Performance under different traffic intensity conditions}

\begin{figure}[htbp]
\begin{centering}
\includegraphics[width=1\columnwidth]{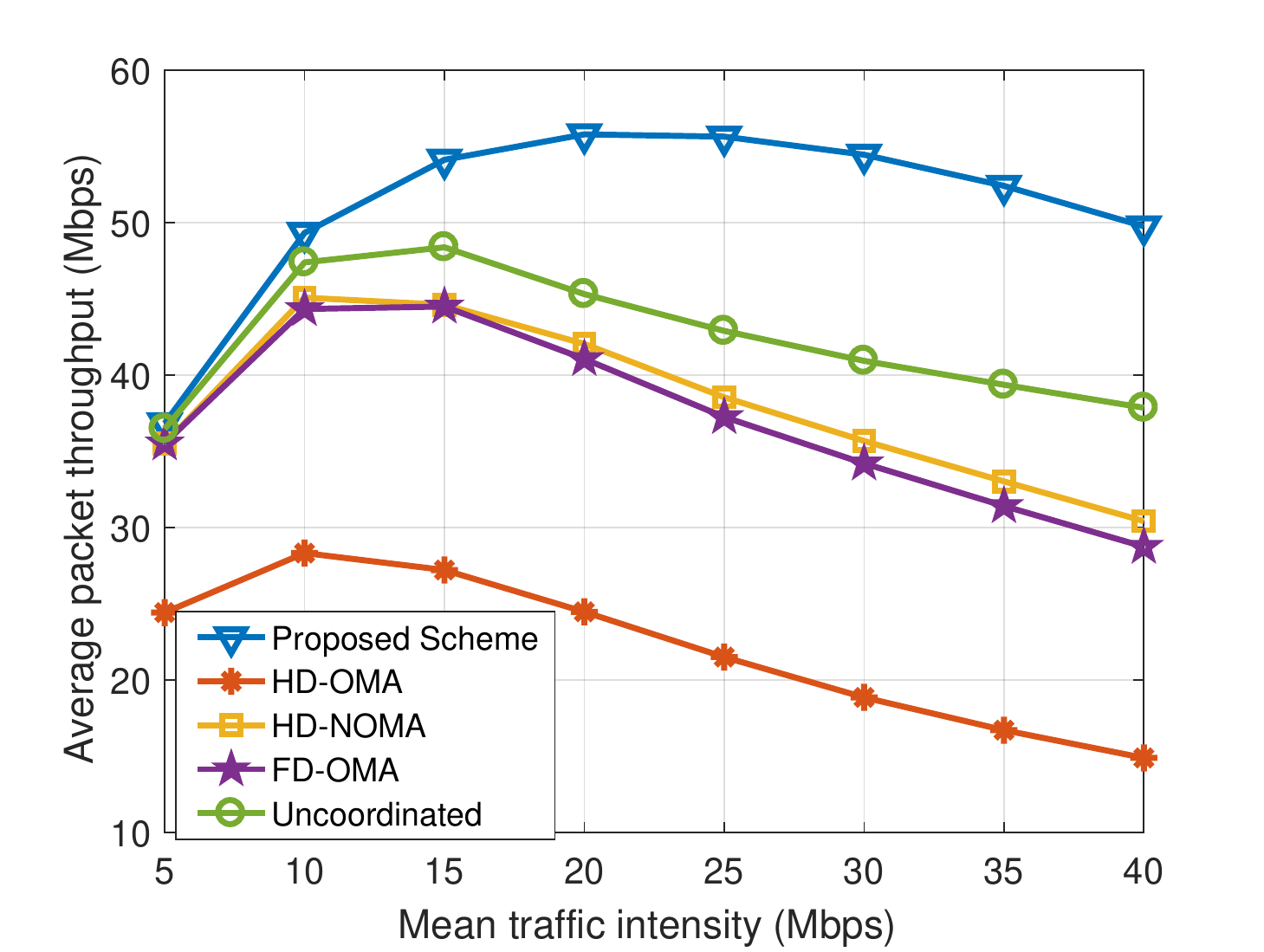}
\par\end{centering}
\caption{Average packet throughput performance for different schemes as the
traffic intensity increases, for a network of $10$ SBSs and an average
of $10$ UEs per SBS. \label{fig:traffic_pckt_thrpt}}
\end{figure}

We start by evaluating the performance of the proposed scheme under
different traffic intensity conditions. An average of $10$ users
per SBS are distributed within the small cell area, with a user mean
packet arrival rate $\lambda_{u}^{\text{total}}=\lambda_{u}^{\text{UL}}+\lambda_{u}^{\text{DL}}$,
and $\lambda_{u}^{\x}=5$ packet/s. Traffic intensity is varied by
changing the mean packet size $1/\mu_{u}^{\x}$ between $50$ kb and
$400$ kb. In Figure \ref{fig:traffic_pckt_thrpt}, the packet throughput
performance is depicted for the different schemes. The packet throughput
is defined as the packet size successfully transmitted to the user
divided by the delay encountered to complete its transmission. In
moderate traffic conditions, the packet throughput increases with
increasing traffic intensity, as the packet size increases and not
much queuing delay is encountered. As the traffic intensity increases,
queuing delay increases, correspondingly decreasing the packet throughput.
Figure \ref{fig:traffic_pckt_thrpt} shows that the proposed scheme
outperforms the baseline schemes in packet throughput performance.
At higher traffic intensity condition, gains of up to $63\%$ and
$73\%$ are observed over HD-NOMA and FD-OMA baseline schemes, respectively.
The performance of the uncoordinated scheme with matching is close
to the proposed scheme at low traffic conditions. However, as the
traffic load increases, the coordination gain reaches up to $31\%$.

\begin{figure}[htbp]
\begin{centering}
\includegraphics[width=1\columnwidth]{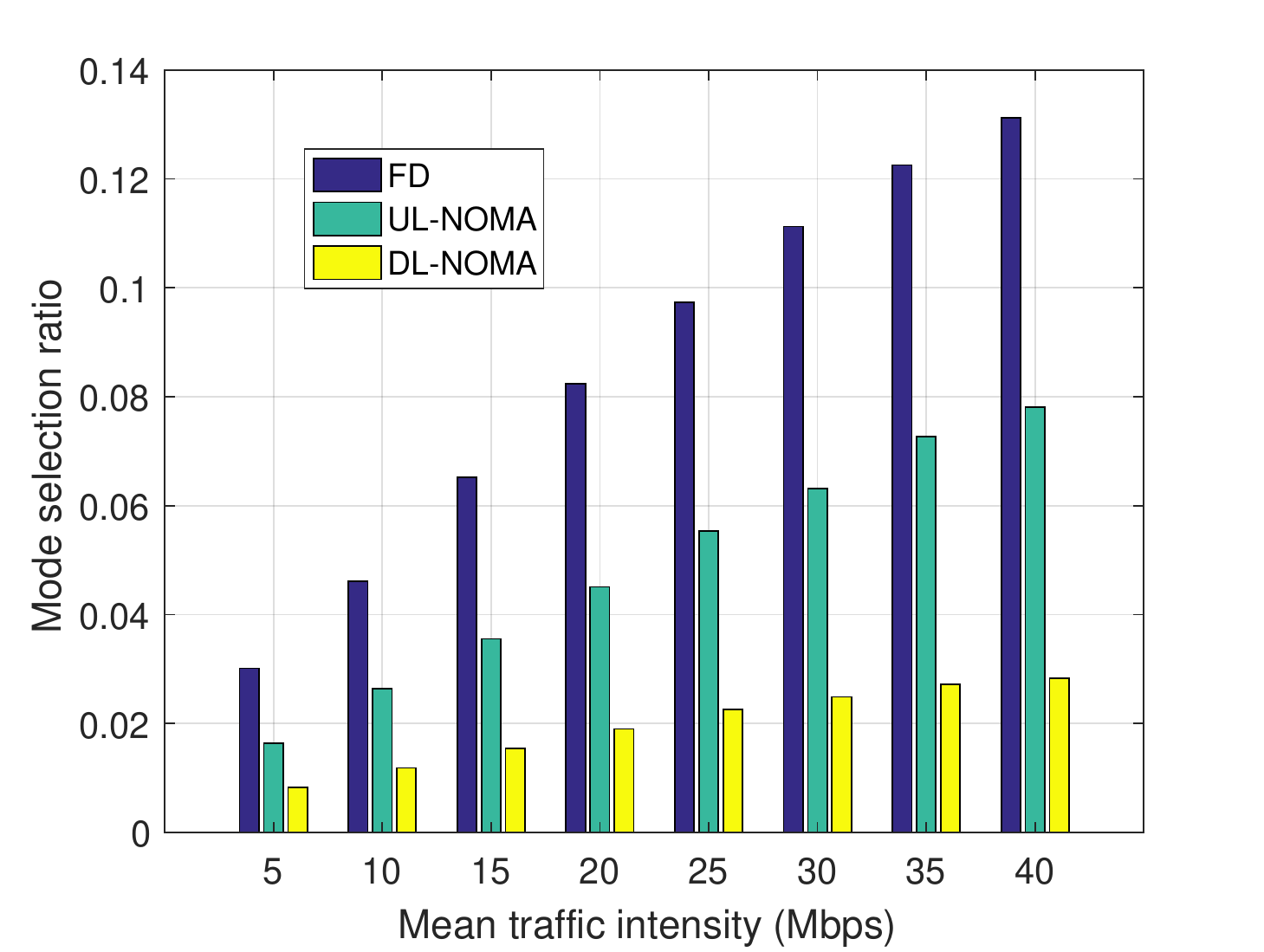}
\par\end{centering}
\caption{FD and NOMA mode selection ratios at different traffic intensity levels,
for a network of $10$ SBSs and an average of $10$ UEs per SBS. Remaining
transmissions are in HD-OMA mode.\label{fig:traffic_percentage}}
\end{figure}

We continue by investigating the effect of traffic intensity on SBS
mode selection. To that end, Figure \ref{fig:traffic_percentage}
provides the ratio of transmissions carried out in FD or NOMA with
respect to all transmissions in different traffic intensity conditions.
Notice here that, although not explicitly represented, HD-OMA was
the operation mode for the remaining of the transmissions.

The results show an increasing rate of operation in both FD and NOMA
as the traffic intensity increases. The use of FD varies from about
$3\%$ at low traffic conditions to $13\%$ at high traffic intensity,
whereas NOMA operation accounts for up to $10\%$ of the transmissions
under high traffic intensity conditions. The rest of transmissions
are in HD-OMA mode. These results clearly indicate that an SBS can
benefit more from scheduling multiple users simultaneously as traffic
intensifies. This is mainly due to the higher request diversity, in
which the chances to have subsets of users that benefit from being
served simultaneously increase.

\begin{figure}[htbp]
\begin{centering}
\includegraphics[width=1\columnwidth]{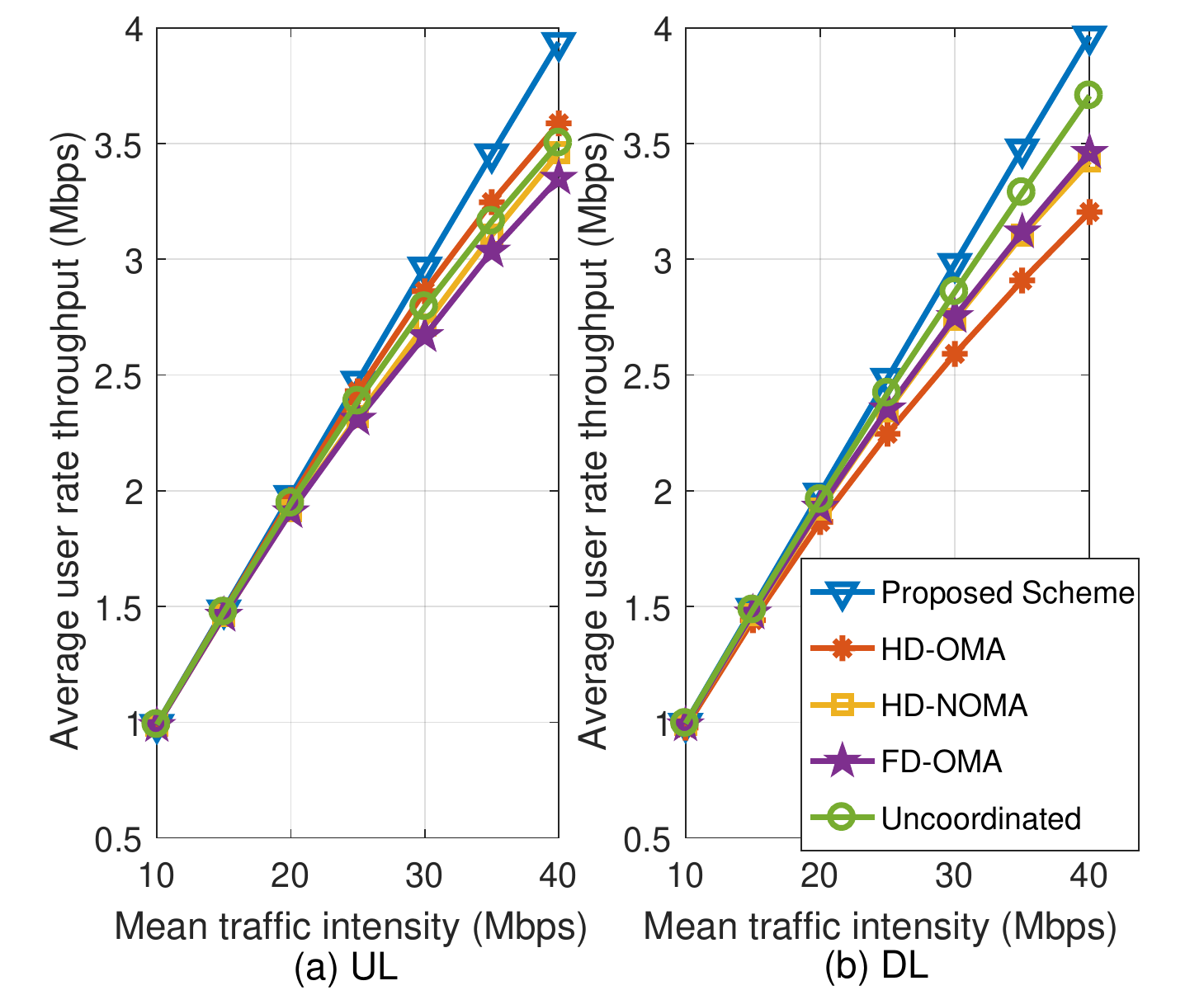}
\par\end{centering}
\caption{Average (a) UL and (b) DL rate throughput performance for different
schemes as the network traffic intensity increases, for a network
of $10$ SBSs and an average of $10$ UEs per SBS.\label{fig:traffic_rate}}
\end{figure}

Next, the impact of traffic intensity on UL/DL user rate throughput
performance is analyzed through Figure \ref{fig:traffic_rate}. A
close up look into Figure \ref{fig:traffic_rate}(a) and Figure \ref{fig:traffic_rate}(b)
shows that results from all baseline schemes fall below HD-OMA for
UL whereas for the DL this scheme is outperformed.

The DL-to-UL interference has a significant impact in the uncoordinated
schemes due to the higher transmitting power of SBSs and lower pathloss
between the SBS and user. In the DL case, the UL-to-DL interference
can be partly avoided within each cell by the pairing thresholds used
in the FD-OMA and HD-NOMA baselines. As our proposed scheme optimizes
the UL and DL power allocation, it achieves UL and DL gains of $9\%$
and $23\%$ over the HD scheme, and of about $12\%$ in both UL and
DL over the HD-NOMA and FD-OMA baseline schemes. Moreover, the coordination
gain over the FD-NOMA matching scheme is even more evident in the
UL ($12\%$) as compared to the DL ($7\%$), which is due to the dominance
of DL-to-UL interference in the uncoordinated scenarios.

\begin{figure}[htbp]
\begin{centering}
\includegraphics[width=1\columnwidth]{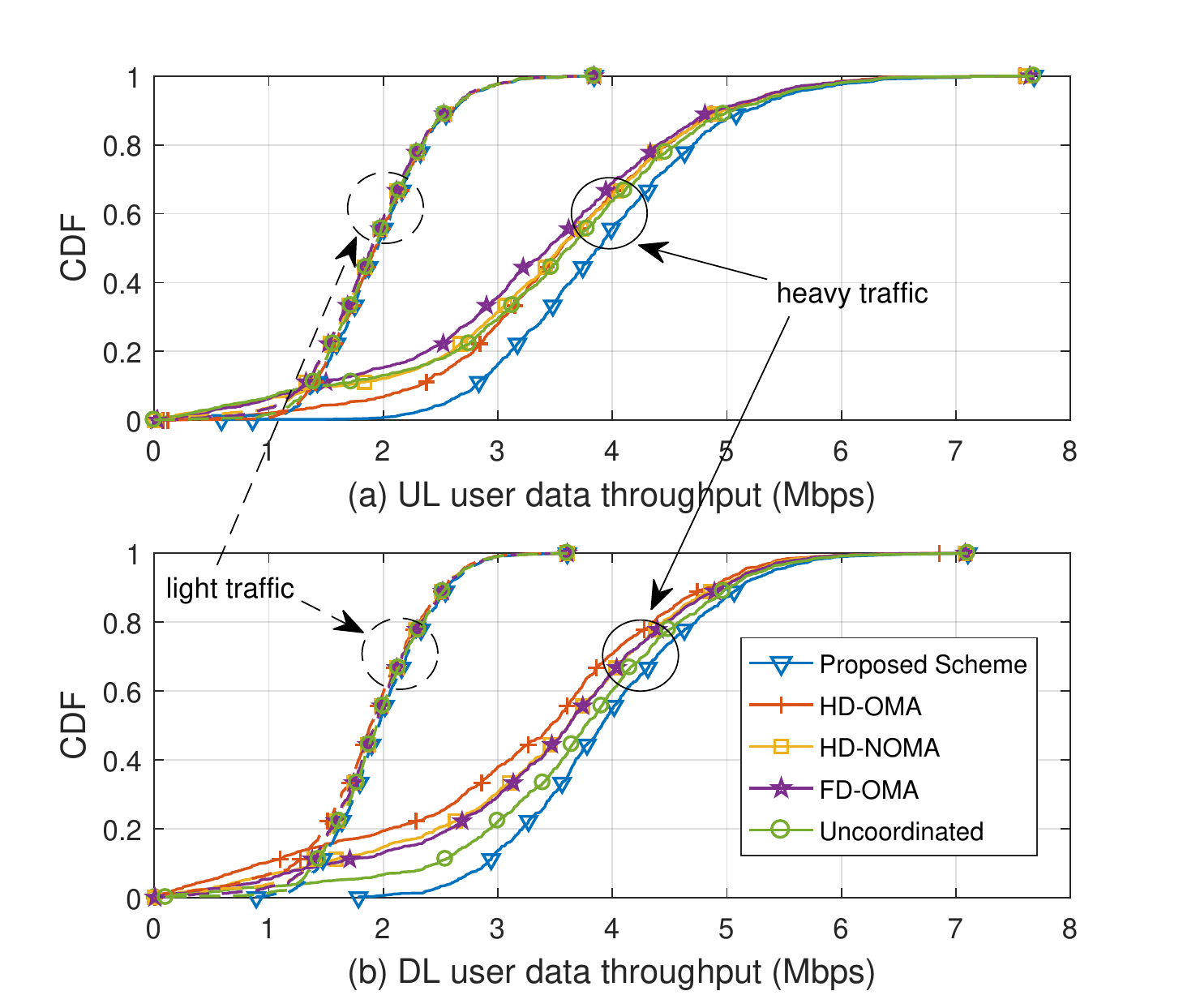}
\par\end{centering}
\caption{CDFs of UL and DL user throughput at light ($20$ Mbps) and heavy
($40$ Mbps) traffic intensity cases for a network of $10$ SBSs and
an average of $10$ UEs per SBS.\label{fig:traffic_cdfs}}
\end{figure}

To further analyze the data rate throughput performance in the UL
and DL, the cumulative distribution functions (CDFs) of the UL and
DL user rate throughput for light and heavy traffic intensity cases
are provided in Figure \ref{fig:traffic_cdfs}. As expected, in the
presence of light traffic no significant gains are achieved with respect
to HD scheme performance; Light traffic does not enable enough diversity
for user scheduling purposes. Accordingly, most of the transmissions
are in HD mode, as was also shown in Figure \ref{fig:traffic_percentage}
for the proposed scheme. Under heavy traffic conditions, the baseline
schemes suffer a degraded cell-edge throughput in the UL due to the
DL-to-UL interference, as discussed above. Figure \ref{fig:traffic_cdfs}
also shows that the proposed scheme power optimization overcomes the
cell-edge throughput degradation and achieves gains of at least $21\%$
and $17\%$ in the UL and DL $10$-th percentile throughput.

\subsection{Performance under different network densities}

\begin{figure}[htbp]
\begin{centering}
\includegraphics[width=1\columnwidth]{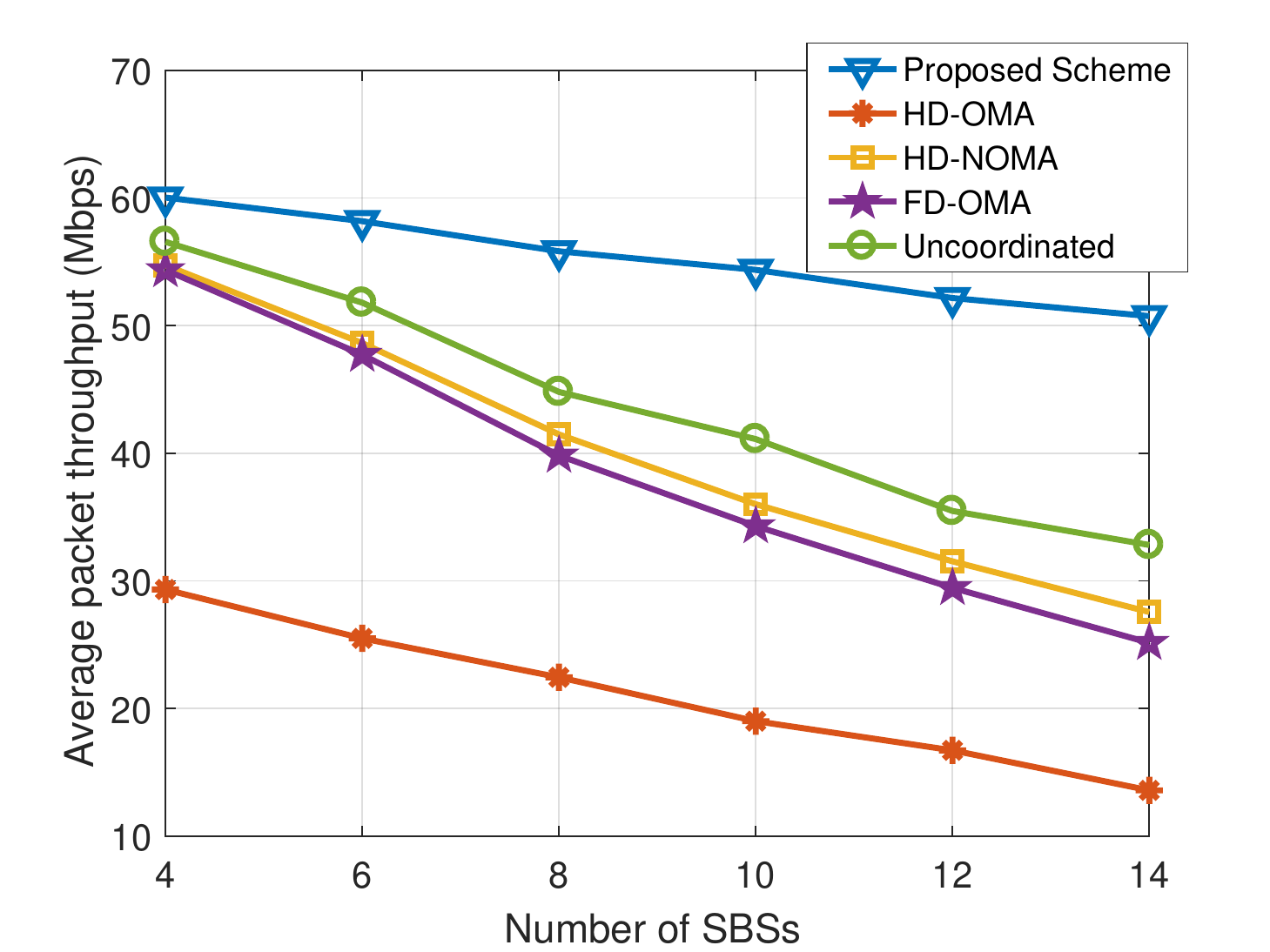}
\par\end{centering}
\caption{Average packet throughput performance for different schemes as the
network density increases for an average of $10$ UEs per SBS and
a mean traffic intensity of $3$ Mbps per user.\label{fig:sbs_pckt_thrpt}}
\end{figure}

We proceed by evaluating the performance of our proposed scheme as
the network density increases. Different network densities will be
simulated by introducing an increased number of SBSs in the system,
while intra-cell user density and traffic influx rates are kept constant.
First, we compare the packet throughput performance against the baseline
schemes. Figure \ref{fig:sbs_pckt_thrpt} shows that, at low network
density, all schemes achieve almost double the packet throughput of
the HD baseline scheme. This is due to multiplexing gain where SBSs
are able to serve more than one user simultaneously and with not much
inter-cell interference to affect the performance. For the same reason,
the uncoordinated scheme performance is close to the proposed scheme,
since coordinating the power to avoid inter-cell interference is not
crucial in this case. On the other hand, as the network density increases,
the coordination gain increases, reaching up to $55\%$ for $14$
SBSs.

\begin{figure}[htbp]
\begin{centering}
\includegraphics[width=1\columnwidth]{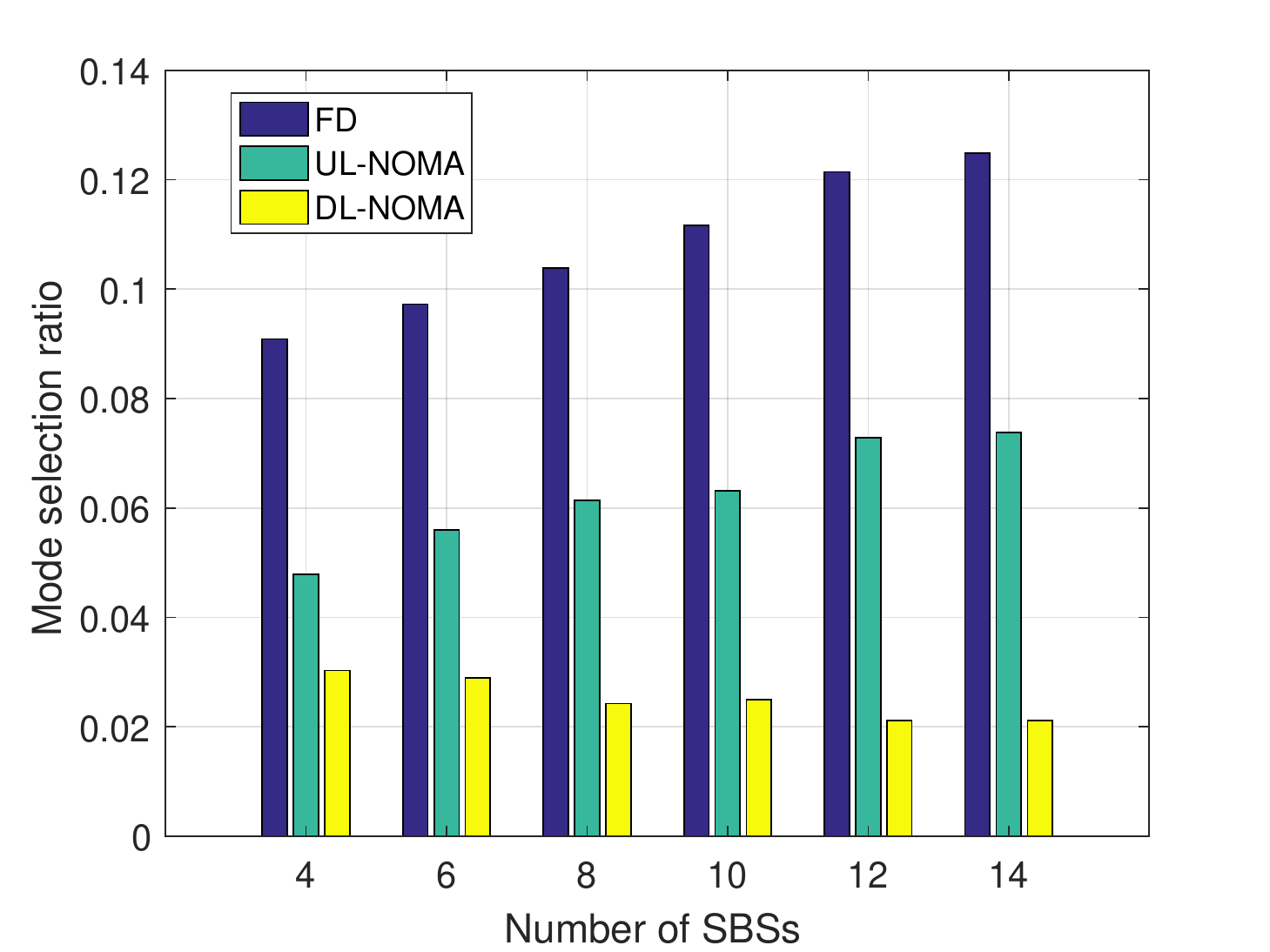}
\par\end{centering}
\caption{FD and NOMA mode selection ratios as the network density increases
for an average of $10$ UEs per SBS and a mean traffic intensity of
$3$ Mbps per user. Remaining transmissions are in HD-OMA mode.\label{fig:sbs_percentage}}
\end{figure}

Next, we investigate the ratio of selecting FD and NOMA modes in the
proposed schemes as the network density varies. As Figure \ref{fig:sbs_percentage}
shows, as the network density increases, the selection of DL NOMA
decreases. The decrease is due to the way power is allocated in DL
NOMA, where users with less channel gains (typically cell-edge users)
have to be allocated higher power as compared to users with higher
channel gains. In that case, user with higher channel gains can decode
and cancel others' data before decoding their own. Therefore, as the
network density increases, inter-cell interference levels increase,
making it is less likely to benefit from DL NOMA. Instead, SBSs schedule
DL users either in HD-OMA or FD modes.

\subsection{Impact of SI cancellation capability}

\begin{figure}[htbp]
\begin{centering}
\includegraphics[width=1\columnwidth]{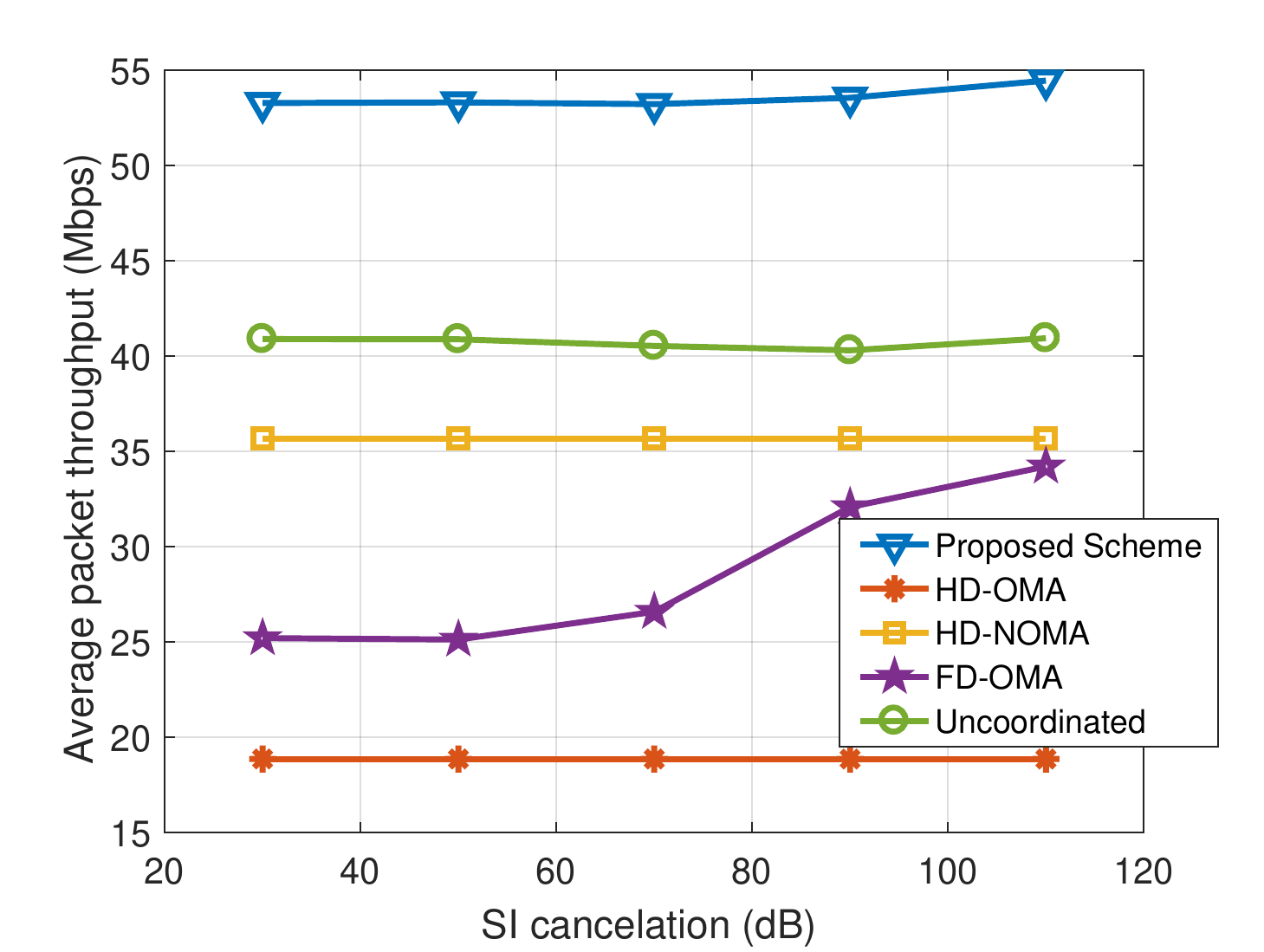}
\par\end{centering}
\caption{Average packet throughput performance for different schemes as the
SBS SI cancellation capability varies, for a network of $10$ SBSs,
an average of $10$ UEs per SBS, and a mean traffic intensity rate
of $3$ Mbps per user.\label{fig:si_pckt_thrpt}}
\end{figure}

Finally, in this subsection, the effect of the SBS SI cancellation
capability on the proposed and baseline schemes is analyzed. Figure
\ref{fig:si_pckt_thrpt} compares the average packet throughput performance
of the different schemes as the SI cancellation varies from $30$
dB to $110$ dB, which is the highest reported capability level. As
Figure \ref{fig:si_pckt_thrpt} shows, the degradation in the proposed
scheme performance due to having lower SI cancellation levels is minor
compared to the FD-OMA baseline scheme. As the proposed scheme optimizes
the mode selection, it can operate more frequently in UL NOMA instead
of FD to serve UL users, so that the high interference from the SBS
DL signal is avoided.

\begin{figure}[htbp]
\begin{centering}
\includegraphics[width=1\columnwidth]{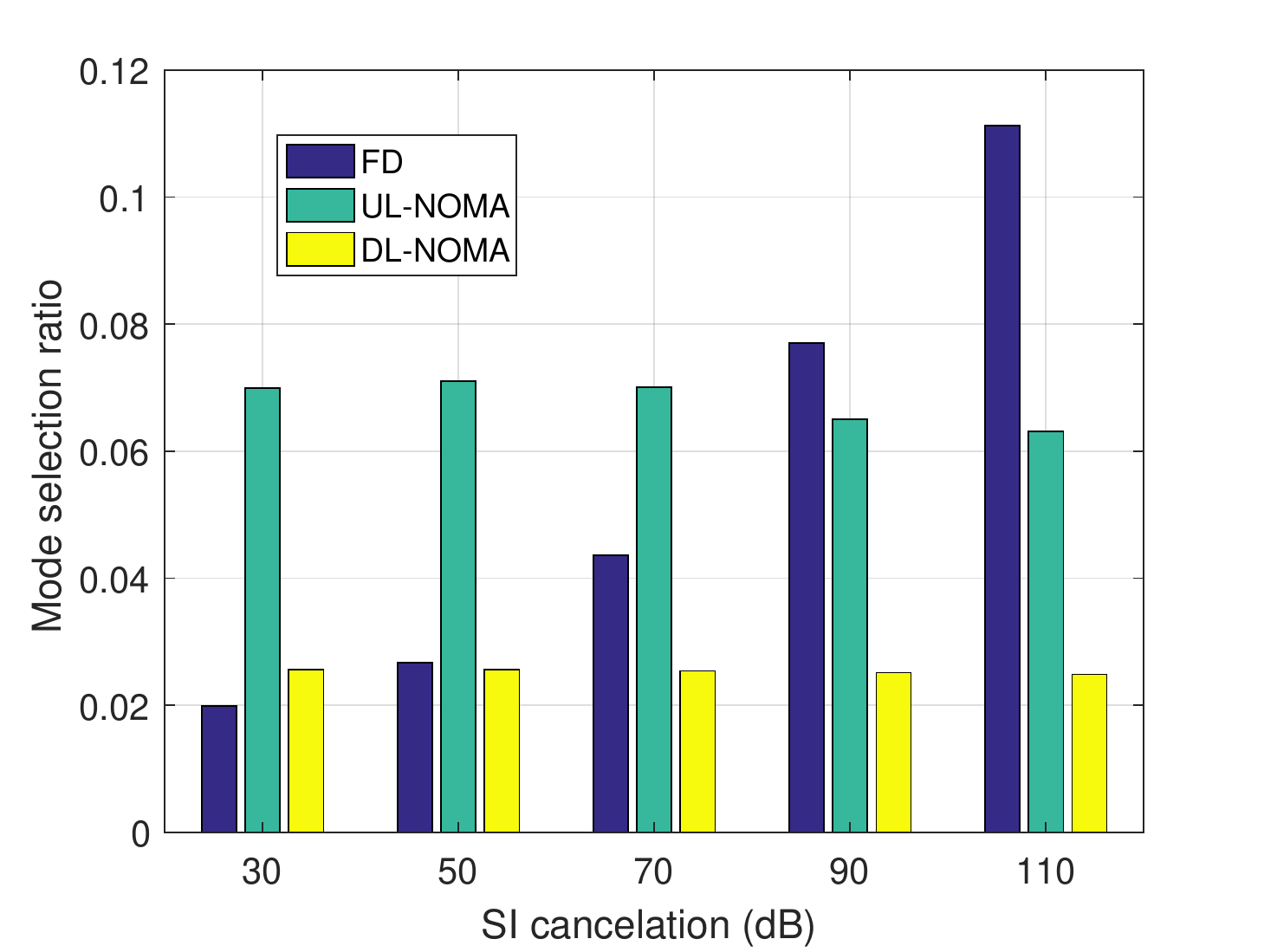}
\par\end{centering}
\caption{FD and NOMA mode selection ratios at different SBS SI cancellation
capability levels, for a network of $10$ SBSs, an average of $10$
UEs per SBS, and a mean traffic intensity rate of $3$ Mbps per user.
Remaining transmissions are in HD-OMA mode.\label{fig:si_percentage}}
\end{figure}

Figure \ref{fig:si_percentage} shows the selection ratio of FD and
NOMA modes in the proposed scheme at different SI cancellation levels.
It can be seen that FD mode is selected more frequently at higher
SI cancellation levels, whereas the UL NOMA ratio decreases correspondingly.
The ratio of DL NOMA selection maintains the same, since SI interference
only affects the UL transmissions.

\section{Conclusions\label{sec:conc}}

In this paper, the problem of mode selection, dynamic user association,
and power optimization has been studied for IBFD and NOMA operating
networks. The problem of time-averaged UL and DL rate maximization
problem under queue stability constraints has been formulated and
solved using Lyapunov framework. A many-to-one matching algorithm
is proposed that locally select which users to serve and in which
transmission mode to operate. Then, a power optimization problem of
the matched users has been formulated and solved iteratively. Simulations
results show significant gains of up to $63\%$ and $73\%$ in UL
and DL packet throughput, and $21\%$ and $17\%$ in UL and DL cell
edge throughput, respectively. Possible future research directions
are to study the problem in ultra-dense network environments as well
as optimizing the network for latency and reliability requirements.
\bibliographystyle{ieeetr}
\bibliography{IEEEabrv,bibfile}

\begin{IEEEbiography}[{\includegraphics[width=1in,height=1.25in,clip,keepaspectratio]{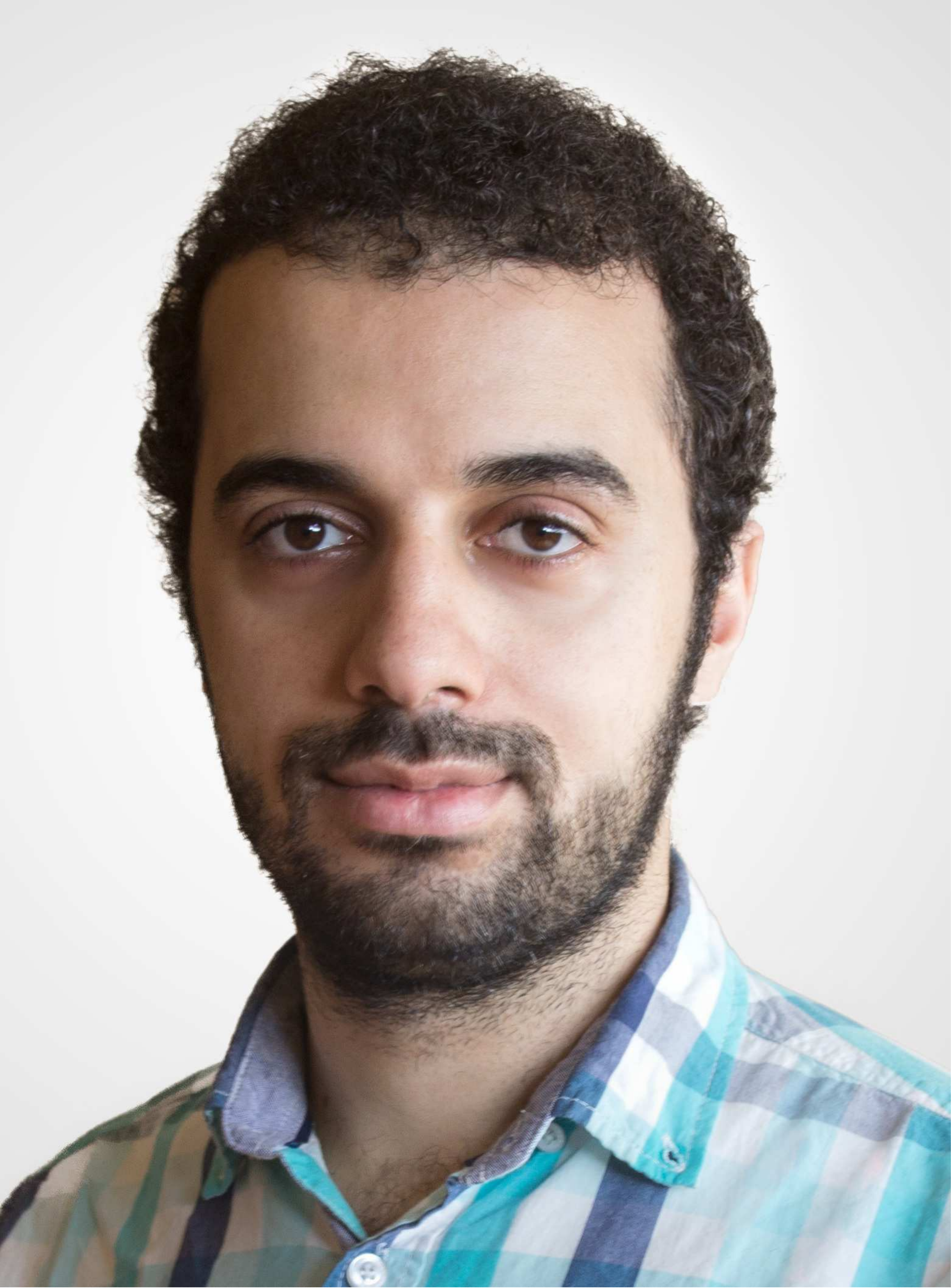}}]{Mohammed S. Elbamby} received his B.Sc. degree (honors) in electronics and communications engineering from the Institute of Aviation Engineering and Technology (IAET), Egypt, in 2010, and his M.Sc. degree in Communications engineering from Cairo University, Egypt, in 2013. After receiving his M.Sc. degree, he joined the Centre for Wireless Communications (CWC) at the University of Oulu, where he is working toward his doctoral degree. His research interests include resource optimization, UL/DL configuration, fog networking, and caching in wireless cellular networks. Mohammed S. Elbamby is a recipient of the Best Student Paper Award from the European Conference on Networks and Communications (EuCNC'2017).
\end{IEEEbiography}

\begin{IEEEbiography}[{\includegraphics[width=1in,clip,keepaspectratio]{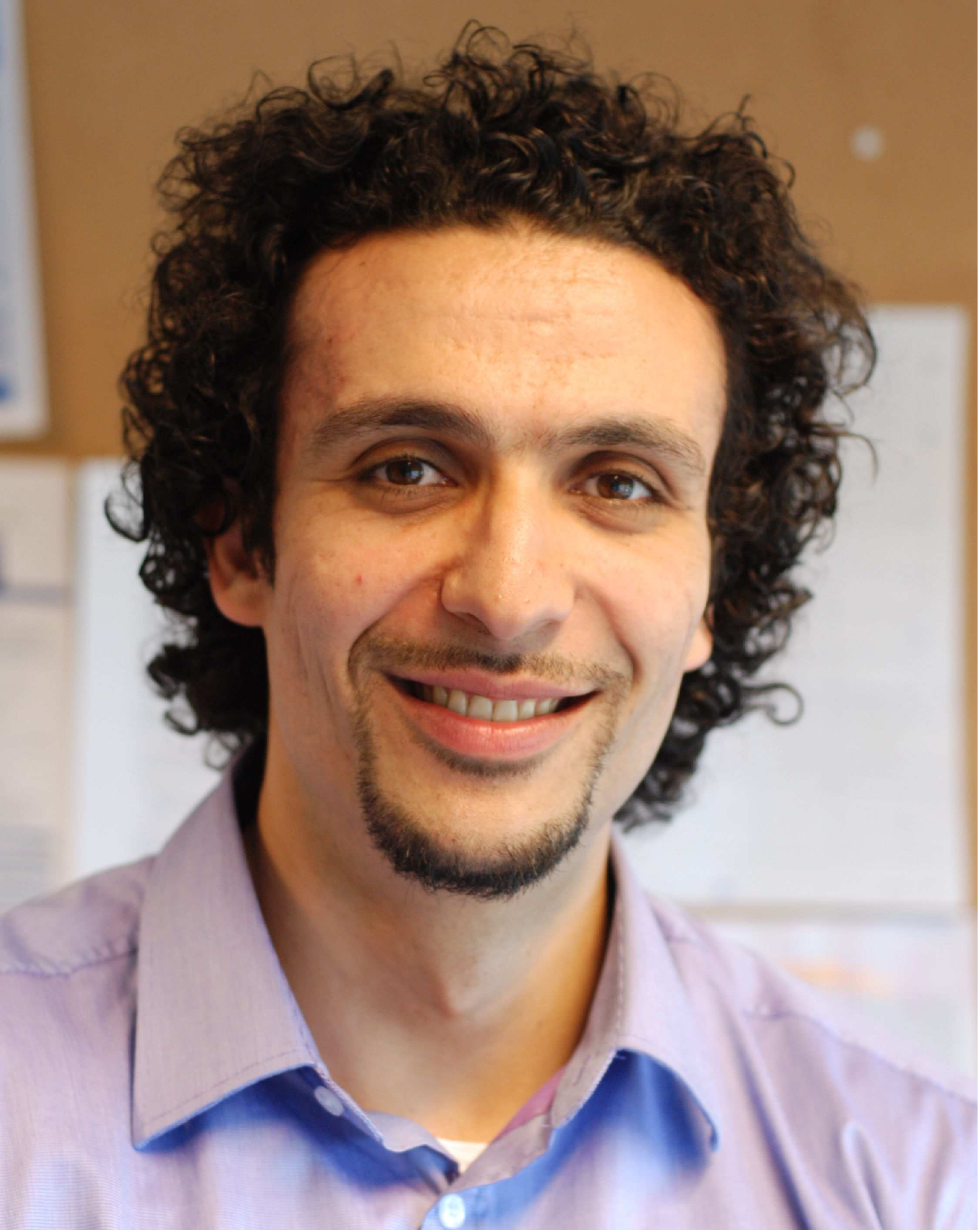}}]{Mehdi Bennis} (Senior Member, IEEE) received his M.Sc. degree in Electrical Engineering jointly from the EPFL, Switzerland and the Eurecom Institute, France in 2002. From 2002 to 2004, he worked as a research engineer at IMRA-EUROPE investigating adaptive equalization algorithms for mobile digital TV. In 2004, he joined the Centre for Wireless Communications (CWC) at the University of Oulu, Finland as a research scientist. In 2008, he was a visiting researcher at the Alcatel-Lucent chair on flexible radio, SUPELEC. He obtained his Ph.D. in December 2009 on spectrum sharing for future mobile cellular systems. Currently Dr. Bennis is an Adjunct Professor at the University of Oulu and Academy of Finland research fellow. His main research interests are in radio resource management, heterogeneous networks, game theory and machine learning in 5G networks and beyond. He has co-authored one book and published more than 100 research papers in international conferences, journals and book chapters. He was the recipient of the prestigious 2015 Fred W. Ellersick Prize from the IEEE Communications Society, the 2016 Best Tutorial Prize from the IEEE Communications Society and the 2017 EURASIP Best paper Award for the Journal of Wireless Communications and Networks. Dr. Bennis serves as an editor for the IEEE Transactions on Wireless Communication.
\end{IEEEbiography}

\begin{IEEEbiography}[{\includegraphics[width=1in,height=1.25in,clip,keepaspectratio]{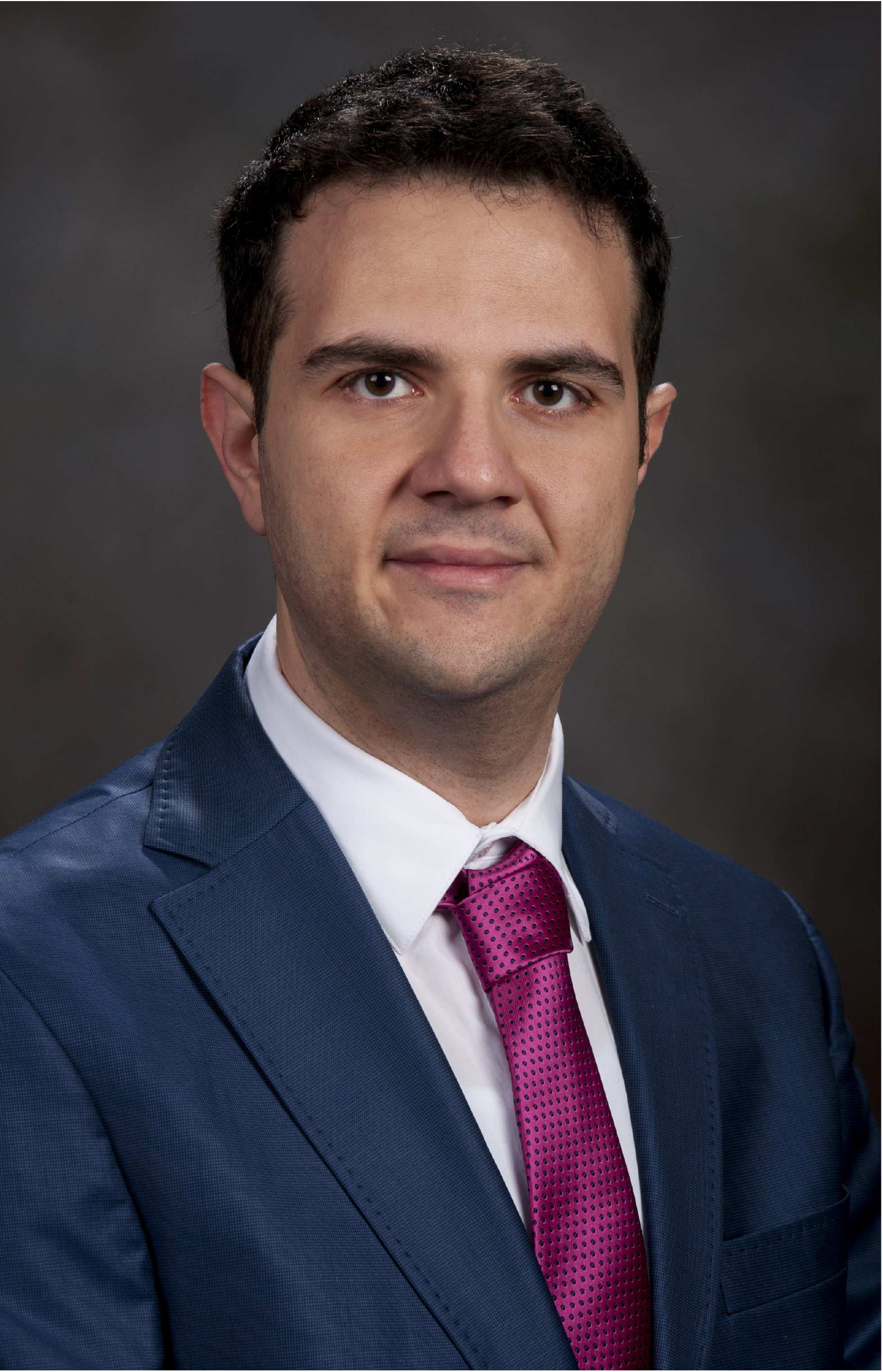}}]{Walid Saad} (S'07, M'10, SM'15) received his Ph.D degree from the University of Oslo in 2010. Currently, he is an Associate Professor at the Department of Electrical and Computer Engineering at Virginia Tech, where he leads the Network Science, Wireless, and Security (NetSciWiS) laboratory, within the Wireless@VT research group. His research interests include wireless networks, game theory, cybersecurity, unmanned aerial vehicles, and cyber-physical systems. Dr. Saad is the recipient of the NSF CAREER award in 2013, the AFOSR summer faculty fellowship in 2014, and the Young Investigator Award from the Office of Naval Research (ONR) in 2015. He was the author/co-author of five conference best paper awards at WiOpt in 2009, ICIMP in 2010, IEEE WCNC in 2012, IEEE PIMRC in 2015, and IEEE SmartGridComm in 2015. He is the recipient of the 2015 Fred W. Ellersick Prize from the IEEE Communications Society. In 2017, Dr. Saad was named College of Engineering Faculty Fellow at Virginia Tech. From 2015 – 2017, Dr. Saad was named the Steven O. Lane Junior Faculty Fellow at Virginia Tech. He currently serves as an editor for the IEEE Transactions on Wireless Communications, IEEE Transactions on Communications, and IEEE Transactions on Information Forensics and Security.
\end{IEEEbiography}

 \begin{IEEEbiography}[{\includegraphics[width=1in,height=1.25in,clip,keepaspectratio]{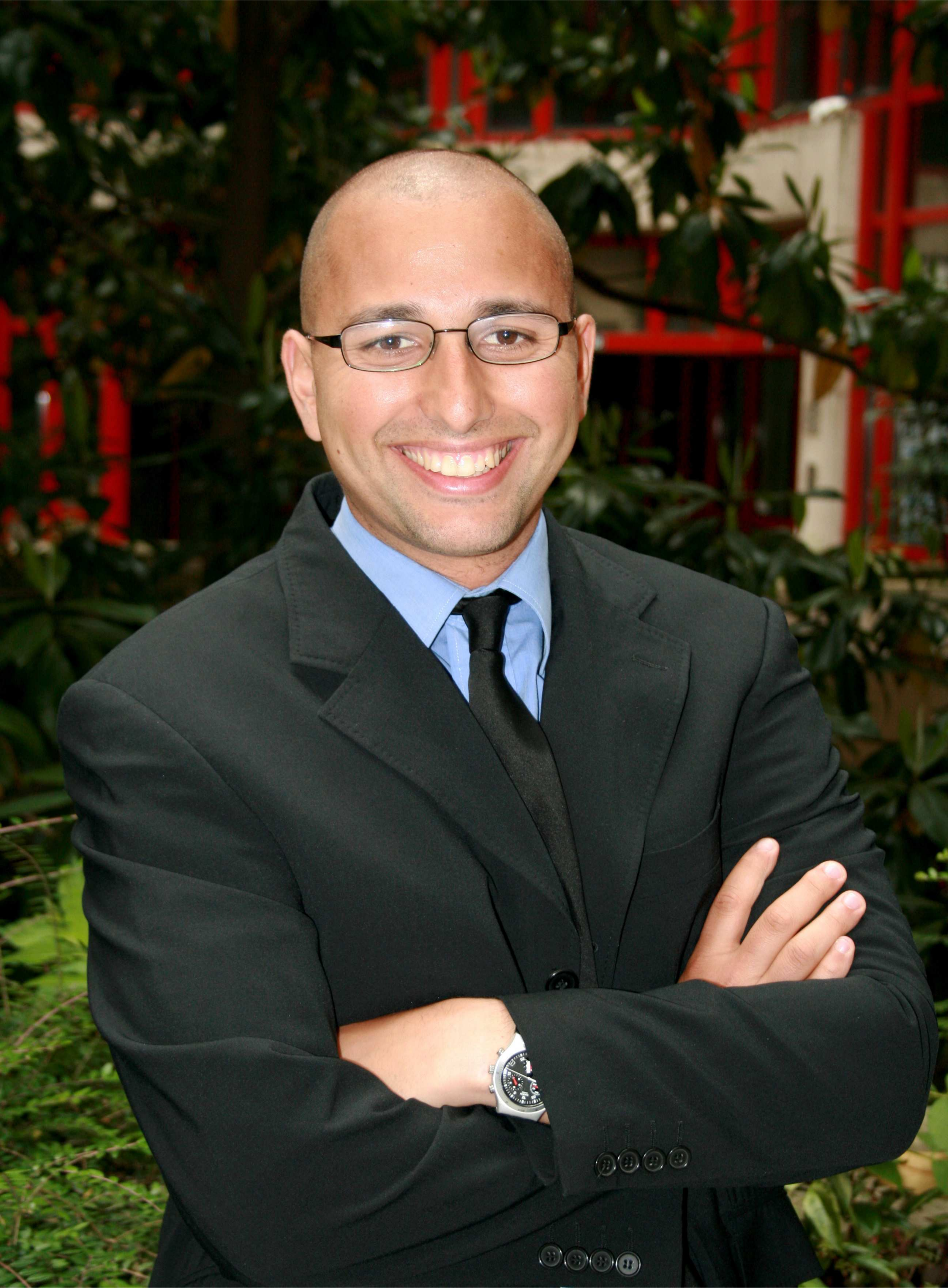}}]{Mérouane Debbah} entered the Ecole Normale Supérieure Paris-Saclay (France) in 1996 where he received his M.Sc and Ph.D. degrees respectively. He worked for Motorola Labs (Saclay, France) from 1999-2002 and the Vienna Research Center for Telecommunications (Vienna, Austria) until 2003. From 2003 to 2007, he joined the Mobile Communications department of the Institut Eurecom (Sophia Antipolis, France) as an Assistant Professor. Since 2007, he is a Full Professor at CentraleSupelec (Gif-sur-Yvette, France). From 2007 to 2014, he was the director of the Alcatel-Lucent Chair on Flexible Radio. Since 2014, he is Vice-President of the Huawei France R\&D center and director of the Mathematical and Algorithmic Sciences Lab. His research interests lie in fundamental mathematics, algorithms, statistics, information \& communication sciences research. He is an Associate Editor in Chief of the journal Random Matrix: Theory and Applications and was an associate and senior area editor for IEEE Transactions on Signal Processing respectively in 2011-2013 and 2013-2014. Mérouane Debbah is a recipient of the ERC grant MORE (Advanced Mathematical Tools for Complex Network Engineering). He is a IEEE Fellow, a WWRF Fellow and a member of the academic senate of Paris-Saclay. He has managed 8 EU projects and more than 24 national and international projects. He received 17 best paper awards, among which the 2007 IEEE GLOBECOM best paper award, the Wi-Opt 2009 best paper award, the 2010 Newcom++ best paper award, the WUN CogCom Best Paper 2012 and 2013 Award, the 2014 WCNC best paper award, the 2015 ICC best paper award, the 2015 IEEE Communications Society Leonard G. Abraham Prize, the 2015 IEEE Communications Society Fred W. Ellersick Prize, the 2016 IEEE Communications Society Best Tutorial paper award, the 2016 European Wireless Best Paper Award and the 2017 Eurasip Best Paper Award as well as the Valuetools 2007, Valuetools 2008, CrownCom2009, Valuetools 2012 and SAM 2014 best student paper awards. He is the recipient of the Mario Boella award in 2005, the IEEE Glavieux Prize Award in 2011 and the Qualcomm Innovation Prize Award in 2012.
  \end{IEEEbiography}

\begin{IEEEbiography}[{\includegraphics[width=1in,height=1.25in,clip,keepaspectratio]{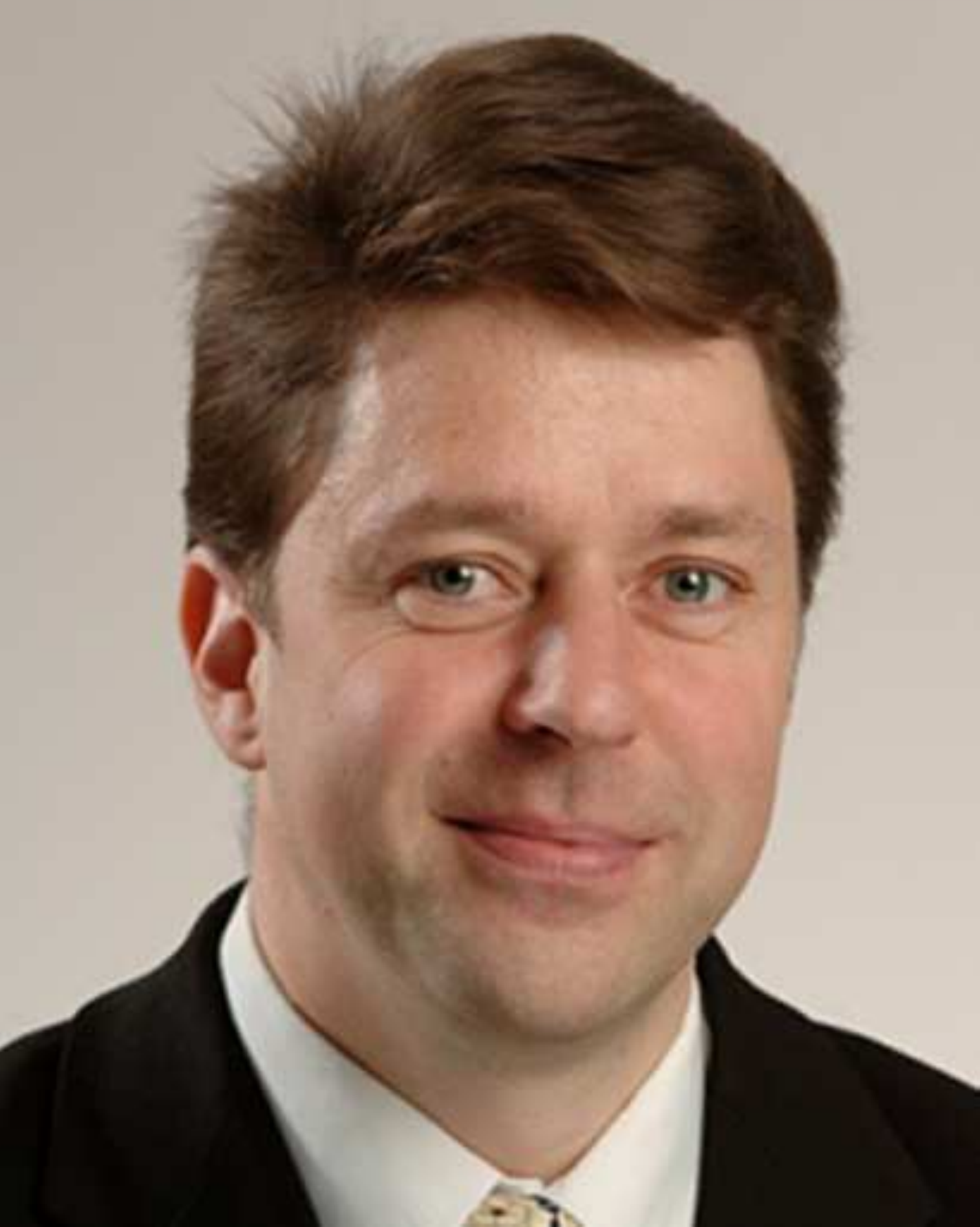}}]{Matti Latva-aho} received the M.Sc., Lic.Tech. and Dr. Tech (Hons.) degrees in Electrical Engineering from the University of Oulu, Finland in 1992, 1996 and 1998, respectively. From 1992 to 1993, he was a Research Engineer at Nokia Mobile Phones, Oulu, Finland after which he joined Centre for Wireless Communications (CWC) at the University of Oulu. Prof. Latva-aho was Director of CWC during the years 1998-2006 and Head of Department for Communication Engineering until August 2014. Currently he is Professor of Digital Transmission Techniques at the University of Oulu. He serves as Academy of Finland Professor in 2017 - 2022. His research interests are related to mobile broadband communication systems and currently his group focuses on 5G systems research. Prof. Latva-aho has published 300+ conference or journal papers in the field of wireless communications. He received Nokia Foundation Award in 2015 for his achievements in mobile communications research.
\end{IEEEbiography}

\end{document}